\documentclass[fleqn,usenatbib]{mnras}

\usepackage{newtxtext,newtxmath}

\usepackage[T1]{fontenc}

\DeclareRobustCommand{\VAN}[3]{#2}
\let\VANthebibliography\thebibliography
\def\thebibliography{\DeclareRobustCommand{\VAN}[3]{##3}\VANthebibliography}

\usepackage{graphicx}	
\usepackage{amsmath}	
\usepackage{orcidlink}
\usepackage{subcaption}
\usepackage{multicol}
\usepackage{units}

\title[multishock scenario for radio relics]{A multishock scenario for the formation of radio relics}

\newcommand{\enzo}{\texttt {ENZO}} 
\newcommand{\CRaTer}{\texttt{CRaTer}}

\newcommand{\music}{\texttt{MUSIC}}
\newcommand{\dd}{\mathrm{d}}

\definecolor{myred}{rgb}{1,0,0} 
\definecolor{myblue}{rgb}{0,0,1}
\definecolor{mygreen}{rgb}{0,1,0}

\author[D. C. Smolinski et al.]{
David C. Smolinski,$^{1}$\thanks{E-mail: david.smolinski@hs.uni-hamburg.de}
Denis Wittor,$^{1,2}$
Franco Vazza \orcidlink{0000-0002-2821-7928}$^{1,2,3}$
and Marcus Brüggen$^{1}$
\\
$^{1}$Hamburger Sternwarte, Universit\"at Hamburg, Gojenbergsweg 112, D-21029 Hamburg, Germany\\
$^{2}$Dipartimento di Fisica e Astronomia, Universita di Bologna, Via Gobetti 93/2, 40122, Bologna, Italy\\
$^{3}$Istituto di Radio Astronomia, INAF, Via Gobetti 101, I-40121 Bologna, Italy
}

\date{Accepted XXX. Received YYY; in original form ZZZ}

\pubyear{2023}

\begin{document}
\label{firstpage}
\pagerange{\pageref{firstpage}--\pageref{lastpage}}
\maketitle

\begin{abstract}
Radio relics are giant sources of diffuse synchrotron radio emission in the outskirts of galaxy clusters that are associated with shocks in the intracluster medium. Still, the origin of relativistic particles that make up relics is not fully understood. For most relics, diffusive shock acceleration (DSA) of thermal electrons is not efficient enough to explain observed radio fluxes.
In this paper, we use a magneto-hydrodynamic simulation of galaxy clusters in combination with Lagrangian tracers to simulate the formation of radio relics. Using a Fokker-Planck solver to compute the energy spectra of relativistic electrons, we determine the synchrotron emission of the relic.
We find that re-acceleration of fossil electrons plays a major role in explaining the synchrotron emission of radio relics. Particles that pass through multiple shocks contribute significantly to the overall luminosity of a radio relic and greatly boosts the effective acceleration efficiency. Furthermore, we find that the assumption that the luminosity of a radio relic can be explained with DSA of thermal electrons leads to an overestimate of the acceleration efficiency by a factor of more than $10^3$. 
\end{abstract}




\section{Introduction}
Astrophysical shock waves are one of the major mechanisms for particle acceleration in the universe and they occur on very different scales, from supernovae scales up to the scales of galaxy clusters. At all these scales, the underlying acceleration mechanism is generally believed to be diffuse shock acceleration \citep[DSA, e.g.][]{1987PhR...154....1B}. DSA is based on Fermi's theory that CR particles can be accelerated by shocks as they are scattered by plasma irregularities between the upstream and the downstream regions \citep{1949PhRv...75.1169F}.\\
Galaxy clusters are astrophysical objects where particle acceleration is observed. Among other components such as Dark Matter $(\sim80-85\%)$ and galaxies, galaxy clusters are essentially composed of the intracluster medium (ICM, $\sim15-20\%$), a $\unit[10^7-10^8]{K}$ hot plasma visible in the X-ray spectrum \citep{2019SSRv..215...16V}.
Mergers between galaxy clusters produce shocks in the ICM which are associated with diffuse synchrotron sources that are called radio relics.
Evidence for a connection between relics and shocks stems from X-ray observations \citep[e.g.]{2014IJMPD..2330007B, 2019SSRv..215...16V}. The CR electrons that are accelerated by DSA 
emit synchrotron radiation  
\citep{1998A&A...332..395E, 1999ApJ...518..603R}. While the connection between shock and radio relic is widely accepted, many details of the acceleration mechanism itself are yet unclear. 

One major issue is the lack of understanding of the acceleration efficiencies at work. The acceleration efficiency measures how much kinetic energy dissipated at the shock goes into the acceleration of the particles, i.e. electrons in case of radio relics. Most of our knowledge of the acceleration efficiencies originates from studies of supernova remnants. The underlying shocks of supernova remnants show very different characteristics than shocks that produce radio relics. Shocks of supernovae have Mach numbers of $M>10^3$. For these strong shocks the acceleration efficiencies and the ratio of CR protons to CR electrons is relatively well known \citep{2011JApA...32..427J, 2012A&A...538A..81M, 2014ApJ...783...91C}. In contrast, the shocks underlying radio relics are weak shocks with Mach numbers of $M\lesssim 3-5$. In this case, the acceleration efficiency is not really known. Current models assume less than a few percent for the acceleration efficiencies \citep{2005ApJ...620...44K, 2013ApJ...764...95K, 2018ApJ...864..105H, 2020MNRAS.495L.112W}. These estimates are based on direct constraints from $\gamma$-ray observations \citep{2010ApJ...717L..71A, 2014ApJ...787...18A, 2016ApJ...819..149A}.\\

Observations of a number of relics suggest that much higher acceleration efficiencies are required to explain the high radio luminosities,
provided that particles are accelerated from the thermal pool.
For some relics, these efficiencies are so high that they violate conservation of energy \citep{2020A&A...634A..64B}. \\
In order to solve this problem, different hypotheses have been proposed, amongst others the re-acceleration of a pre-existing population of CR electrons \citep{2011ApJ...734...18K, 2012ApJ...756...97K, 2013MNRAS.435.1061P}. In this case, the CR electrons are not accelerated from the thermal pool, but from an already mildly relativistic seed population. Previous encounters with shock waves or Active Galactic Nuclei (AGN) have been proposed as a source for such a population of fossil CR electrons \citep[e.g.][]{2014ApJ...785....1B, 2018ApJ...865...24D}. This hypothesis is supported by observations of a connection between a radio relic and an AGN \citep{2017NatAs...1E...5V}. However, the AGN scenario is only feasible if the AGN jets are predominantly composed of electrons and positrons to keep the number of CR protons low enough to explain their non-detection in the $\gamma$-rays \citep[][]{2015MNRAS.451.2198V, 2019ApJ...871..195A}.

Here, we show that the Multishock Scenario (MSS) can explain the high acceleration efficiency of radio relics \citep{2022MNRAS.509.1160I}. 
In the MSS, it is assumed that a fraction of the cosmic-rays that form the relic have undergone previous episodes of shock acceleration.
The shocks that the cosmic rays have previously encountered could be accretion shocks, or shocks from previous mergers or other violent processes in the ICM.
MSS has been studied before in various contexts \citep[e.g.][]{1997PASA...14..251M,2020JKAS...53...59K,2021A&A...647A..94S,2022MNRAS.509.1160I}. 
Using an analytical approach, \citet{1997PASA...14..251M} and \citet{2021A&A...647A..94S} showed that the CR spectra produced by MSS depend on many factors, such as the distance between two shocks but, most prominently, the Mach number of the shocks.
Using a cosmological simulation, \citet{2022MNRAS.509.1160I} have shown that the MSS can explain the high luminosities of radio relics.
Yet, it is still unclear if the MSS can also explain the unrealistically high acceleration efficiencies that are inferred for several radio relics assuming acceleration from the thermal pool.
Moreover, it has been shown that the shock obliquity, i.e. the angle between the shock normal and the upstream magnetic field, plays a crucial role in the shock acceleration of thermal electrons \citep{2014ApJ...794..153G,2014ApJ...797...47G}. It is still unknown how the shock obliquity affects the MSS.

In this paper, we compute the acceleration efficiencies that would be inferred for a radio relic that has been produced by MSS. To this end, we use a cosmological simulation to study the evolution of CR electrons during the formation of a massive galaxy cluster. Using Lagrangian tracer particles, we follow the spectral evolution of CR electrons that form a radio relic. We compute the relic's emission with and without the re-acceleration of CR electrons. To understand the effect of re-acceleration, we compare the acceleration efficiencies inferred from the radio luminosity to the model efficiencies. Finally, we study the role of the shock obliquity in the MSS. \\

This paper is organized as follows: In Sec.~\ref{sec:simulations}, we describe the numerical set-up used for the galaxy cluster simulation, as well as the tools and methods used for our analysis.
In Sec.~\ref{sec:properties}, we evaluate the properties of the selected relic. In Sec.~\ref{sec:evo_e_spec}, we present the results of the radio spectra, from this we computed in Sec.~\ref{sec:radio_emission_result} the synchrotron emission of the relic. In Sec.~\ref{sec:acc_eff_result}, we study the acceleration efficiencies one would get from observation versus the given model. Sec.~\ref{sec:conclusion} summarizes our results.

\section{Simulations} \label{sec:simulations}

\subsection{Cosmological simulations in \enzo}\label{sec:cosmsim}

In order to produce realistic simulations of the formation of a massive galaxy cluster, we use the magneto-hydrodynamical code \enzo\; \citep{2014ApJS..211...19B}. We take a simulation from the San Pedro-cluster catalogue \citep{2021MNRAS.506..396W}.  
The simulation starts at a redshift of $z=30$. 

On the root grid, our simulation covers $\unit[140]{Mpc}\; h^{-1}$ sampled with $256^3$ cells and $256^3$ dark matter particles, this corresponds to a resolution of $\Delta x\approx\unit[0.547]{Mpc}\; h^{-1}$. 
Furthermore, we add five levels of nested grids centered at the final location of the galaxy cluster of interest.
We use the \music-code to initialize the five nested grids \citep{2011MNRAS.415.2101H}.

On the fifth level, our simulation covers $\unit[6.56]{Mpc}\; h^{-1}$ sampled with $384$ cells, corresponding to a resolution of $ \Delta x\approx\unit[17.09]{kpc}\; h^{-1}$.  At redshift 1, we add one additional layer of adaptive-mesh refinement (AMR) in \enzo. As the AMR criterion, we use the \textit{MustRefineRegion}-criterion, implemented in \enzo. The sixth level covers a volume of $\unit[2.73]{Mpc}\;h^{-1}$, that is sampled with a uniform spatial resolution of $\Delta x\approx\unit[8.55]{kpc}\;h^{-1}$. The AMR region is centered on a galaxy cluster with a total mass $M\approx\unit[1.5\cdot 10^{15}]{M_\odot}$ that experiences several mergers. 

For the MHD solver, we use the local Lax-Friedrichs (LLF) Riemann solver to compute the fluxes in the piece-wise linear method (PLM). We initialized a magnetic field of $\unit[10^{-7}]{G}$ in each direction at the start of the simulation ($z=30$). We use Dedner-cleaning in order to produce physically correct magnetic fields \citep[][]{2002JCoPh.175..645D,2018SSRv..214..122D}. In this paper, we use the following cosmological parameters: $h=0.6766$, $\Omega_\Lambda=0.69$, $\Omega_{\rm m}=0.31$ and $\Omega_{\rm b}=0.05$, in agreement with the latest results from the Planck Collaboration \citep{2020A&A...641A...1P}.

\subsection{Lagrangian tracer particles simulated with \CRaTer}\label{sec:Lagrangiantracers}

In post-processing, we track the cosmic rays using Lagrangian tracers. To this end, we use the Lagrangian code \CRaTer\ that computes the spatial and temporal evolution of cosmic rays. For more details regarding \CRaTer, we refer the reader to \citet{2016Galax...4...71W, 2017MNRAS.464.4448W, 2017MNRAS.471.3212W}. In the beginning, we initialise the tracer particles such that they follow the  distribution of the ICM and we injected a total of $\sim 2.1\cdot 10^6$ tracers at a redshift of $z=1$. Furthermore, at runtime, \CRaTer \ injects new particles following the continuous infall of matter from the boundaries of the computational domain. The result is that at redshift $z = 0$, the cluster is sampled with $\sim 17.7\cdot 10^6$ tracers.
The corresponding gas mass that is associated with a single tracer is then $m_{\mathrm{trac}}=\unit[4.7\cdot 10^7]{M_\odot}$.
In order to assign the various gridded physical quantities, such as magnetic fields or densities, to the tracer particles, we use a cloud in cell (CIC) interpolation with correction factors on the velocity, see \citep{2017PhDT.......222W}. 
\newline
In order to detect when the tracer particles experience a shock, we use a temperature-jump-based shock finder that is based on the Rankine-Hugoniot jump conditions \citep[][]{2017PhDT.......222W, 2017MNRAS.464.4448W}. The Mach number for each shock is then given by \cite{2020MNRAS.495L.112W}:

\begin{equation}
    M=\sqrt{\frac{4}{5}\frac{T_{\mathrm{new}}}{T_{\mathrm{old}}}\frac{\rho_{\mathrm{new}}}{\rho_{\mathrm{old}}}+\frac{1}{5}} .
\end{equation}
The calculation is applied to a tracer at two consecutive timesteps. The subscripts "old" and "new" here indicate the different values before and after the shock.
The tracers also keep track of the pre-shock obliquity, which is the angle between the vector of the local magnetic field and the shock normal, i.e.

\begin{equation}
    \Theta_\mathrm{pre}=\arccos{\left(\frac{\Delta \textbf{v} \cdot \textbf{B}_\mathrm{pre}}{|\Delta \textbf{v}||\textbf{B}_\mathrm{pre}|}\right)} ,
\end{equation}
where $\textbf{B}_\mathrm{pre}$ is the pre-shock magnetic field. The velocity jump between pre- and post-shock gas is given by $\Delta \textbf{v}=\textbf{v}_\mathrm{post}-\textbf{v}_\mathrm{pre}$.

\subsection{Evolution of the electron spectra}\label{sec:evo_of_e}

In order to model the CR electron spectra, we use the  ROGER{\footnote{https://github.com/FrancoVazza/JULIA/tree/master/ROGER}} code \citep{2021A&A...653A..23V, 2023A&A...669A..50V}. With ROGER, we solve the time-dependent diffusion-loss equation for the population of relativistic electrons represented by the Lagrangian tracers, under the assumption of negligible CR diffusion. ROGER uses the standard \citet{1970JCoPh...6....1C} finite difference scheme, coded in parallel using the programming language Julia. 
We used $N_\mathrm{b}=57$ momentum bins equally spaced in $\log(p)$ in the $p_{\mathrm{min}}\leq p \leq p_{\mathrm{max}}$ momentum range, with $P=\gamma\cdot m_e \cdot v$ and $p=P/(m_e\cdot c)$ is the normalized momentum of electrons. We chose $p_{\mathrm{min}}=2$, $p_{\mathrm{max}}=10^6$, \textbf{$p_\mathrm{cut}\approx8\cdot10^5$}and $d\log(p)=0.1$. Considering the Fokker-Planck equation without injection and escape terms, we can calculate the spectrum of our relativistic electrons. Here, $N(p)$ represents the number density of relativistic electrons as a function of momentum for each tracer, and obeys:

\begin{equation}\label{eq:Fokker-Planck_in_Chang-Cooper}
    \frac{\partial N}{\partial t}=\frac{\partial}{\partial p}\Bigl[ N \Bigl(\Bigl|\frac{p}{\tau_{\mathrm{rad}}}\Bigr|+\Bigl|\frac{p}{\tau_\mathrm{c}}\Bigr|+\frac{p}{\tau_{\mathrm{adv}}}-\Bigl|\frac{p}{\tau_{\mathrm{acc}}}\Bigr|\Bigr)\Bigr] .
\end{equation}
A numerical solution can be found via \citep[cf.][]{1970JCoPh...6....1C}:
\begin{equation}\label{numerical_int}
    N(p,t+dt)=\frac{N(p,t)/dt+N(p+dp,t+dt)\dot{p}}{1/dt+\dot{p}/dp}+Q_{\rm inj} .
\end{equation}
As described in \citet{2023A&A...669A..50V}, the solver subcycles between the time steps of the \CRaTer \ outputs, in order to resolve the rapid cooling at high relativistic momenta.

Here, we give a short summary of the gain and loss terms of Eq.~\ref{eq:Fokker-Planck_in_Chang-Cooper}, and we refer to \citet{2021A&A...653A..23V, 2023A&A...669A..50V} for more details. 
$\tau_{\mathrm{rad}}$, $\tau_{\mathrm{c}}$ and $\tau_{\mathrm{adv}}$ are the time scales for radiative losses, Coulomb losses and adiabatic expansion, respectively. We neglect bremsstrahlung because it is significantly less important.
Moreover, CR electrons can gain energy via Fermi-I-type acceleration, i.e., diffusive shock acceleration. In our case, we neglected Fermi-II-type acceleration since it is believed to play a minor role in relics. 

For the injection of the particles, we use a power-law momentum distribution, i.e.

\begin{equation}
    Q_{\mathrm{inj}}(p)=K_{\mathrm{inj}}\cdot p^{-\delta_{\mathrm{inj}}}\cdot \Bigl(1-\frac{p}{p_{\mathrm{cut}}}\Bigr)^{\delta_{\mathrm{inj}}-2} .
    \label{eq:qinj}
\end{equation}
$Q_{\mathrm{inj}}(p)$ denotes the momentum spectrum of injected electrons, i.e. the number of injected CR electrons per normalized momentum $p$, where $K_{\mathrm{inj}}$ is a unitless normalisation factor, $\delta_{\rm inj}=2(M^2+1)/(M^2-1)$ is the slope of the input momentum spectrum and $p_{\rm cut}$ is the cutoff momentum. Since we are using a power-law spectrum, a cutoff momentum is necessary to limit the momentum range in the range of interest. Above this limit, the cooling time scale gets shorter than the acceleration time scale. For all plausible choices the acceleration time scale is determined by the energy-dependent diffusion coefficient, yielding acceleration time scales that are many orders of magnitude smaller than the cooling time scales. The power-law assumption is very useful for the reason that it allows us to simplify the injection of new particles. A new population of particles can simply be added to Eq. \ref{numerical_int} without integrating a source term.
Following this, the total cosmic ray energy per tracer particle, $E_\mathrm{CR}$, is computed by integrating the product of the power law $Q_{\mathrm{inj}}(p)$ and $T(p)=m_{\mathrm{e}} c^2\cdot \bigl(\sqrt{1+p^2}-1\bigr)$
\begin{equation}\label{eq:ecrint}
    E_{\mathrm{CR}}=\int_{p_{\mathrm{inj}}}^{p_{\mathrm{cut}}}Q_{\mathrm{inj}}(p)T(p)dp.
\end{equation}
The integration yields into an expression for $E_{\mathrm{CR}}$:
\begin{equation}\label{ECRexpression}
    E_{\mathrm{CR}}=\frac{K_{\mathrm{inj}}m_ec^2}{\delta_{\mathrm{inj}}-1}\Bigl[ \frac{B_x}{2}\Bigl( \frac{\delta_{\mathrm{inj}}-2}{2}, \frac{3-\delta_{\mathrm{inj}}}{2}\Bigr)+p_{\mathrm{cut}}^{1-\delta_{\mathrm{inj}}}\Bigl(\sqrt{1+p_{\mathrm{cut}}^2}-1\Bigr)\Bigr] .
\end{equation}
Here, $B_x(a,b)$ is the incomplete Bessel function and $x=1/(1+p_{\mathrm{cut}}^2)$  \citep[see][]{2013MNRAS.435.1061P, 2023A&A...669A..50V}.

In order to determine the normalisation factor $K_\mathrm{inj}$ in Eq. \ref{eq:qinj}, we equated a fraction $\eta$ of the kinetic energy flux dissipated at the shock multiplied with the tracer area element and the shock crossing time, $t_\mathrm{cross}=\dd x_t/v_d$, with the total CR energy of each tracers, $E_\mathrm{CR}$.  Here $v_d$ is the downstream velocity of the gas. I.e. we demand that

\begin{equation}
     {1\over 2}\eta(M,\Theta)\cdot (\rho_{\mathrm{u}} \cdot v_{\mathrm{s}}^3\cdot \dd x_{\mathrm{t}}^2) \cdot t_{\mathrm{cross}}= E_\mathrm{CR}
    \label{eq:energyatshock}
\end{equation}

Here $\rho_{\mathrm{u}}$ is the pre-shock gas density, $v_{\mathrm{s}}$ is the shock velocity and $\dd x_{\mathrm{t}}^2$ is the surface associated with the tracers. 
$\dd x_t$ is calculated for each tracer by taking the cubic root of:
\begin{equation}
    \dd x_t^3=\dd x^3/n_\mathrm{tracer}
\end{equation}
where $\dd x^3$ denotes the volume initially associated with every tracer and $n_\mathrm{tracer}$ denotes the number of tracers in every cell.
$\eta(M,\Theta)$ is the CR acceleration efficiency and is given by
\begin{equation}\label{eq:obliquity_cut}
 \eta(M,\Theta)=\frac{1}{2} \bigl( \tanh\bigl(\frac{\Theta-\Theta_c}{\delta}\bigr)+1\bigr)\cdot \tilde{\eta}(M)\cdot \xi_{\mathrm{e}}. 
\end{equation}
Here, the efficiency, $\eta$, depends on the pre-shock obliquity, $\Theta$, the electron to proton ratio, $\xi_{\mathrm{e}}$, and the Mach number, $M$, since $\tilde{\eta}$ depends on $M$. $\tilde{\eta}$ denotes the acceleration efficiency that remains by factoring out the angular dependence.
In our model,  the acceleration efficiency $\eta(M, \Theta)$ depends on the pre-shock obliquity and therefore on the local magnetic field topology. Following \citet{2023MNRAS.519..548B}, we used $\delta=\pi/18$, and we set $\Theta_c=\pi/3$ and $\Theta_c=\pi/4$ in the cases of re-acceleration and acceleration, respectively. Since we are using electrons, the sign of $\Theta$ and $\Theta_c$ is switched in comparison to \citet{2023MNRAS.519..548B}. For $\Tilde{\eta}(M)$, we used the polynomial approximation presented in \citet{2013ApJ...764...95K}. For the weak ICM shocks, the electron-to-proton ratio is very uncertain \citep{2023A&A...669A..50V}. Hence, we use  $\xi_e=(m_p/m_e)^{(1-\delta_{\rm inj})/2}$ \citep{2013MNRAS.435.1061P}. \\

Ignoring the cut-off in the momentum spectrum, the fraction of CRe to thermal electrons is roughly given by

\begin{equation}
\frac{N_{\rm CR}}{N_{\rm th}}\approx 
\frac{K_{\rm{inj}}\int p^{-\delta_{\rm{inj}}}dp}{M_{\rm tracer}/\mu_em_p} \approx \frac{\eta \mu_e m_p v_s^2 p_{\rm min}^{1-\delta_{\rm inj}}}{ m_e c^2\beta}\approx 10^{-5},
\label{eq:ncr}
\end{equation}
where $\mu_e$ is the mean mass per (thermal) electron, $M_{\rm tracer}$ the mass of a single tracer particle and $\beta$ the value of the incomplete Bessel function from Eq. 7, which we take to be 0.1. Moreover, in the last part of Eq.~\ref{eq:ncr}, we assumed $v_s=1000$ km/s, $\delta_{\rm{inj}}=3.3$ and $\eta=10^{-5}$.\
\\
ROGER also models the process of re-acceleration by shocks, which plays a major role for our work.
Upon diffusive shock re-acceleration, the CR momentum spectrum, $N_0(p)$, evolves as 

\begin{equation}\label{eq:Reaccevolution}
    N(p)=(\delta_{\mathrm{inj}}+2)\cdot p^{-\delta_{\mathrm{inj}}}\int_{p_{\min}}^{p_{\mathrm{max}}}N_0(x)x^{\delta_{\mathrm{inj}}+1}\dd x
\end{equation}
which subsequently ages until the next re-acceleration event \citep[cf.][]{2005ApJ...627..733M, 2011ApJ...734...18K, 2012ApJ...756...97K}.

\subsection{Synchrotron emission}\label{sec:synchroemission}

We use two different aproaches for the calculation of the synchrotron radiation. For the first approach, we use the function given in \citet{2007MNRAS.375...77H} (cf. Eq. 32 in the referenced paper). This allows us to compute the radio emission analytically for each tracer:

\begin{equation}\label{eq:HB07_without_prefactors}
    \frac{dP}{d\nu}=C\cdot A \cdot n_{\mathrm{e,-4}} \cdot \xi_{\mathrm{e}} \cdot \nu_{1.4}^{\delta_\mathrm{inj}/2}\cdot T_{\mathrm{d}}^{3/2}\frac{B^{1+\delta_\mathrm{inj}/2}}{B_{\mathrm{CMB}}^2+B^2}\cdot\eta(M) .
\end{equation}
Here $C$ is a normalisation factor, $A$ is the assumed surface area of the shock that produces the relic, $n_{\mathrm{e,-4}}$ is the number density of the electrons in units of $10^{-4}$ cm$^{-3}$, $\xi_{\mathrm{e}}$ is the fraction of CR electrons to protons, $\nu_{1.4}$ is the observing frequency in units of 1.4~GHz, $T_{\mathrm{d}}$ is the downstream temperature, $\delta_\mathrm{inj}$ the spectral index, $B$ is the local magnetic field at the position of the tracer and $B_\mathrm{CMB}$ is the magnetic field strength of the cosmic microwave background. The magnetic field of the cosmic microwave background is given by $B_{\mathrm{CMB}}=\unit[3.2\cdot(1+z)^2]{\mu G}$. Eq. \ref{eq:HB07_without_prefactors} is computed using the quantities recorded by the tracers.

The first approach to the calculation of the radio emission does not include re-acceleration of fossil electrons. Moreover, \citet{2007MNRAS.375...77H} derived Eq. \ref{eq:HB07_without_prefactors} under the assumption of a quasi-stationary balance between the energy gains and losses of CR electrons. However, the resolution of our simulation is below the electron cooling length of the ICM, that is $10-100 \ \mathrm{kpc}$ \citep[e.g.][]{2007MNRAS.375...77H,2012ApJ...756...97K}. Hence, the assumption of a quasi-stationary balance is not applicable, and the spectral evolution of CR electrons must be carefully modelled. Nevertheless, we use the approach for the calculation of the radio emission in our algorithm that clusters the tracers belonging to the same relic, see Sec.~\ref{sec:HOP}.

In the second approach, we compute the synchrotron power as the convolution of the aged CR spectrum $N(\gamma)$ and the modified Bessel function $F\Bigl(\frac{\nu}{\nu_\mathrm{c}}\Bigr)$: 

\begin{equation}\label{eq:synchro_power}
    P(\nu)=\frac{2\pi\sqrt{3}e^2\nu_\mathrm{L}}{c}\int_{\gamma_1}^{\gamma_2}d\gamma N(\gamma) F\Bigl(\frac{\nu}{\nu_\mathrm{c}}\Bigr) .
\end{equation}

Here, $\nu_\mathrm{L}$ is the Lamor frequency, the characteristic frequency is $\nu_\mathrm{c} =(3/2)\gamma^2\nu_\mathrm{L}$. The synchrotron function $F(\nu/\nu_\mathrm{c})$ is given by 

\begin{equation}\label{synchro_kernel}
    F(x)=x\int_x^{\infty}dx'K_{5/3}(x') .
\end{equation}
using $x = \nu/\nu_\mathrm{c}$ and the modified Bessel-function $K_{5/3}$.
To compute Eq.~\ref{eq:synchro_power}, we used the analytical approximation derived by \citet{2014MNRAS.442..979F}:

\begin{equation}
    P(\nu)=P_1\cdot F_p(x,\Lambda) ,
\end{equation}
with $P_1=\pi\sqrt{3}e^2\nu_\mathrm{L} \gamma_1^{-p+1} C/c$ and the parametric function  $F_p(x,\Lambda)$ where $x$ is the dimensionless frequency $x=\nu/\nu_1$ and $\Lambda$ is the ratio of the Lorentz-factors $\Lambda=\gamma_2/\gamma_1$. We are using a form of the equation that leads to small errors of $\sim 10\ \%$  \citep[compare fourth form in Sec.~3.4][]{2014MNRAS.442..979F}. They describe the parametric function $F_p$ by two fitting functions that depend on $x$:

\begin{equation}
    F_p(x,\Lambda)=
    \begin{cases}
    F_p(x)-\Lambda^{-p+1}F_p(x/\Lambda^2), & \text{for } x < x_c\\
    \sqrt{\frac{\pi}{2}}\Lambda^{-p+2}x^{-1/2}\exp\left(\frac{-x}{\Lambda^2}\right)\left[1+a_p\left(\frac{x}{\Lambda^2}\right)^{-1}\right] & \text{for } x \geq x_c .
    \end{cases}
\end{equation}

The fitting formula for $F_p(x)$ is given by 
\begin{equation}
    \begin{split}
    F_p(x)\approx &\kappa_p x^{1/3} \exp(a_1x^2+a_2x+a_3x^{2/3})\\
    &+C_p x^{-(p-1)/2}[1-\exp(b_1x^2)]^{p/5+1/2} .
    \end{split}
\end{equation}

For the values of the constants, we refer the reader to \citet{2014MNRAS.442..979F}. For CR spectra steeper than $p\geq6$, the synchrotron emission is negligible and set to $P(\nu)=0$ in this model.

\subsection{HOP halo finder for grouping the tracer particles}
\label{sec:HOP}

In order to find a relic in our simulation, we search for a spatially connected radio emitting structures in the outskirts of the simulated cluster. Specifically, we search for structures that have a high radio power and have the typical shape of a relic. To this end we use a HOP-algorithm, which is a group-finding algorithm used to find structures in N-body simulations \citep{1998ApJ...498..137E}.\\

The HOP-algorithm uses the local density of the individual particles in order to jump to the neighbour with the next highest density. This happens until there is no neighbour with a higher density than the particle itself. All particles that are connected via jumps to this particle are considered a group. We use the HOP-algorithm included in the yt astro analysis toolkit \citep{yt,yt.astro.analysis}.\\

In our case, instead of the density we use the radio luminosity. For the computation of the radio luminosity, we used Eq.~\ref{eq:HB07_without_prefactors}, which allows for a fast and efficient computation. However, Eq.~\ref{eq:HB07_without_prefactors} neglects the contribution of cosmic-rays, that already underwent some amount of cooling, to the radio emission. Yet, these particles must be accounted for, when using Eq. \ref{eq:synchro_power} to compute the relic emission.
Hence, we assign all particles to the relic that are located inside a cell that contains shocked particles.

\section{Results}

\subsection{Properties of the selected radio relic}
\label{sec:properties}
\begin{figure*}
    \centering
    \begin{subfigure}{0.49\textwidth}
    \includegraphics[width=0.9\linewidth, scale=1.0]{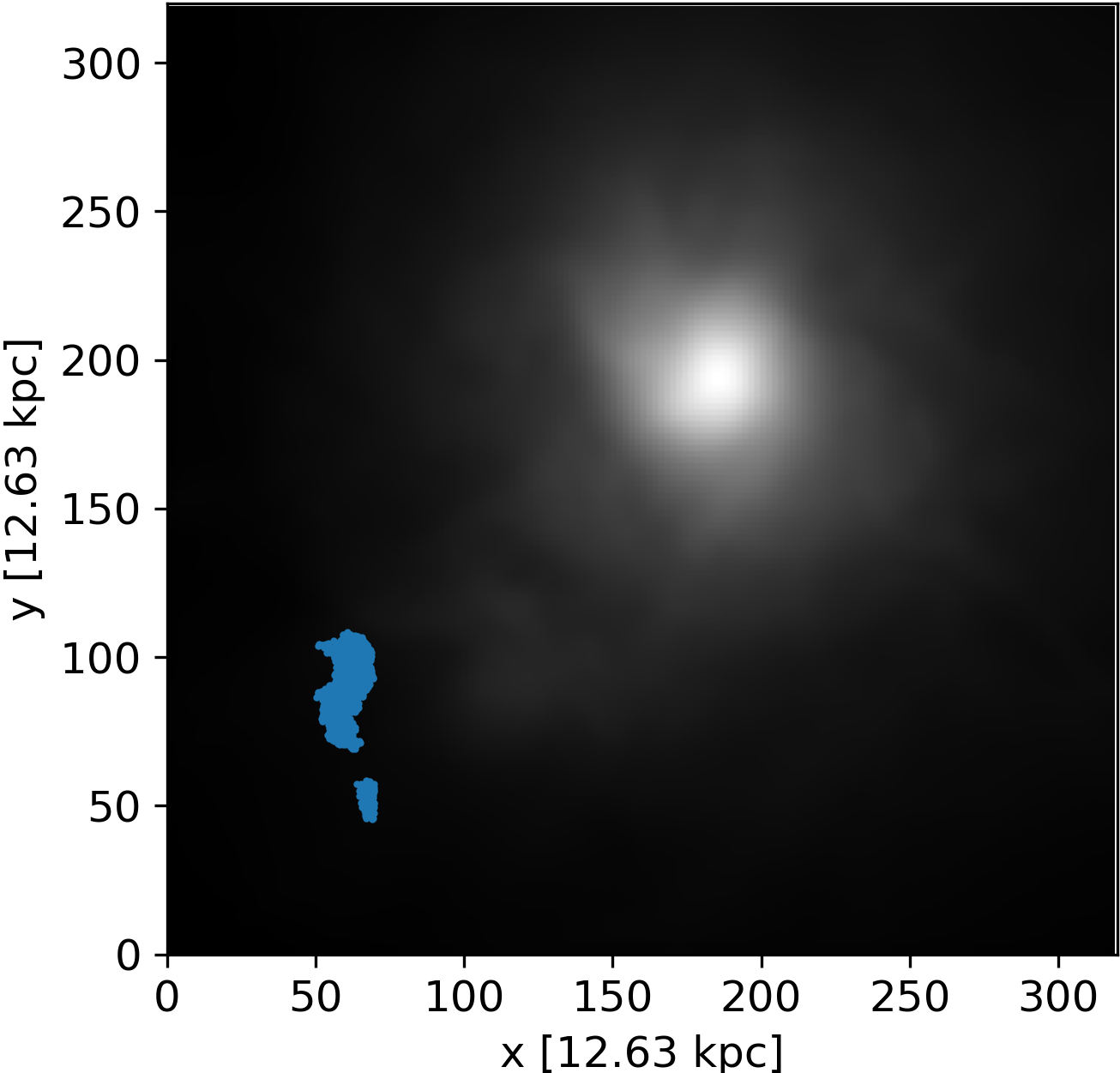}
    \label{subfig:evo_a}
    \caption{}
    \end{subfigure}
    \begin{subfigure}{0.49\textwidth}
    \includegraphics[width=0.9\linewidth, scale=1.0]{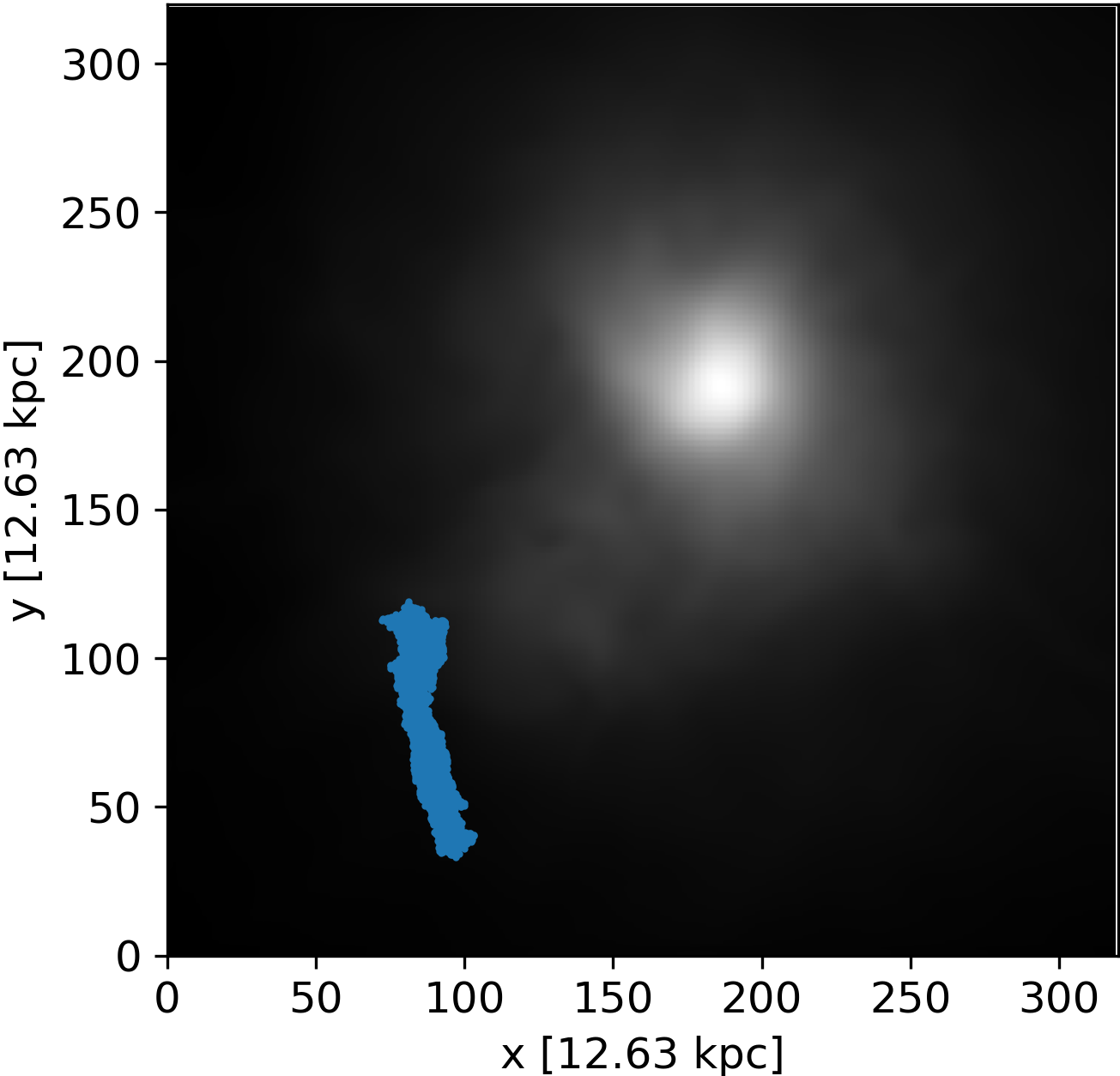}
    \label{subfig:evo_b}
    \caption{}
    \end{subfigure}
    \begin{subfigure}{0.49\textwidth}
    \includegraphics[width=0.9\linewidth, scale=1.0]{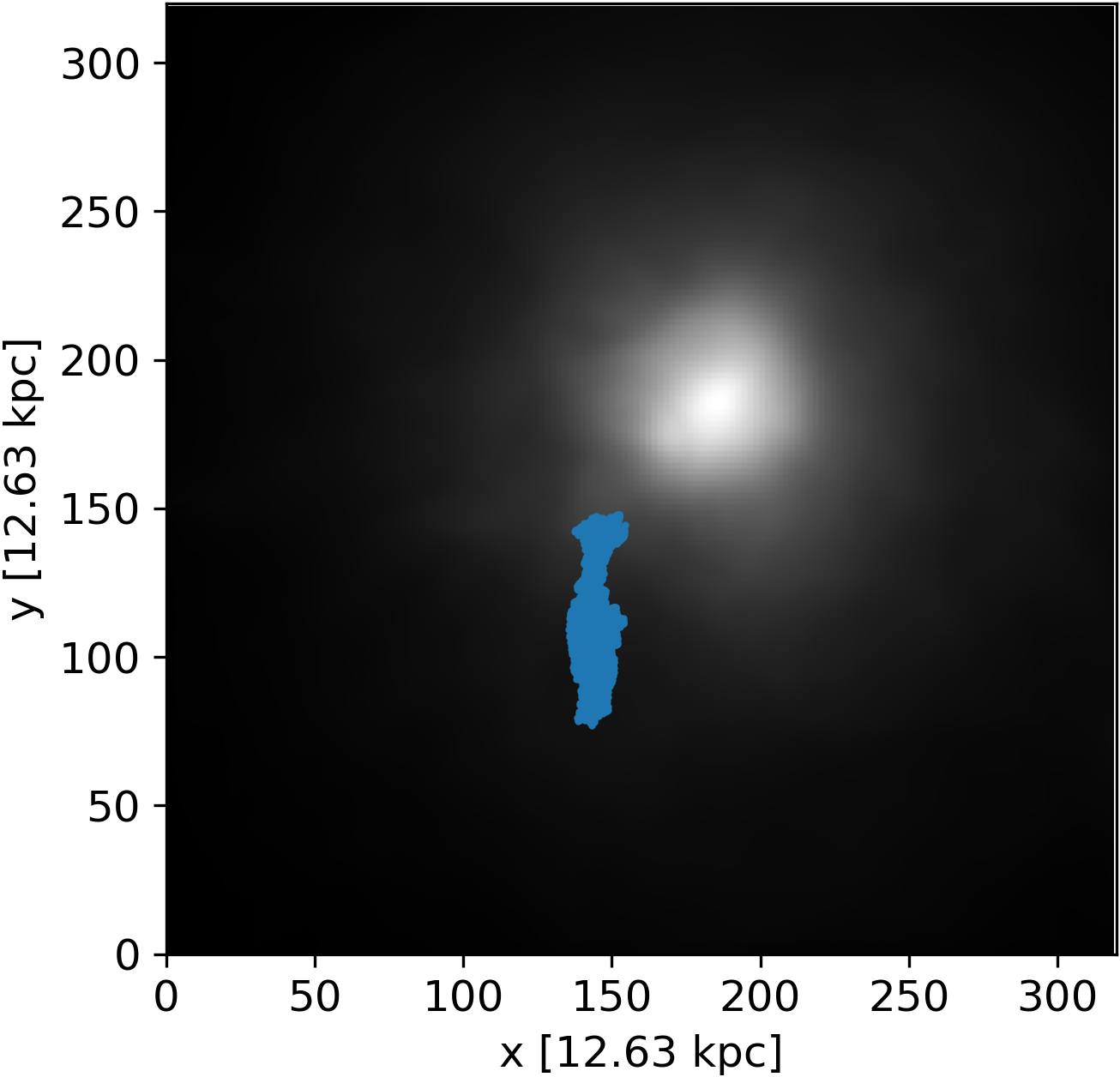}
    \label{subfig:evo_c}
    \caption{}
    \end{subfigure}
    \begin{subfigure}{0.49\textwidth}
    \includegraphics[width=0.9\linewidth, scale=1.0]{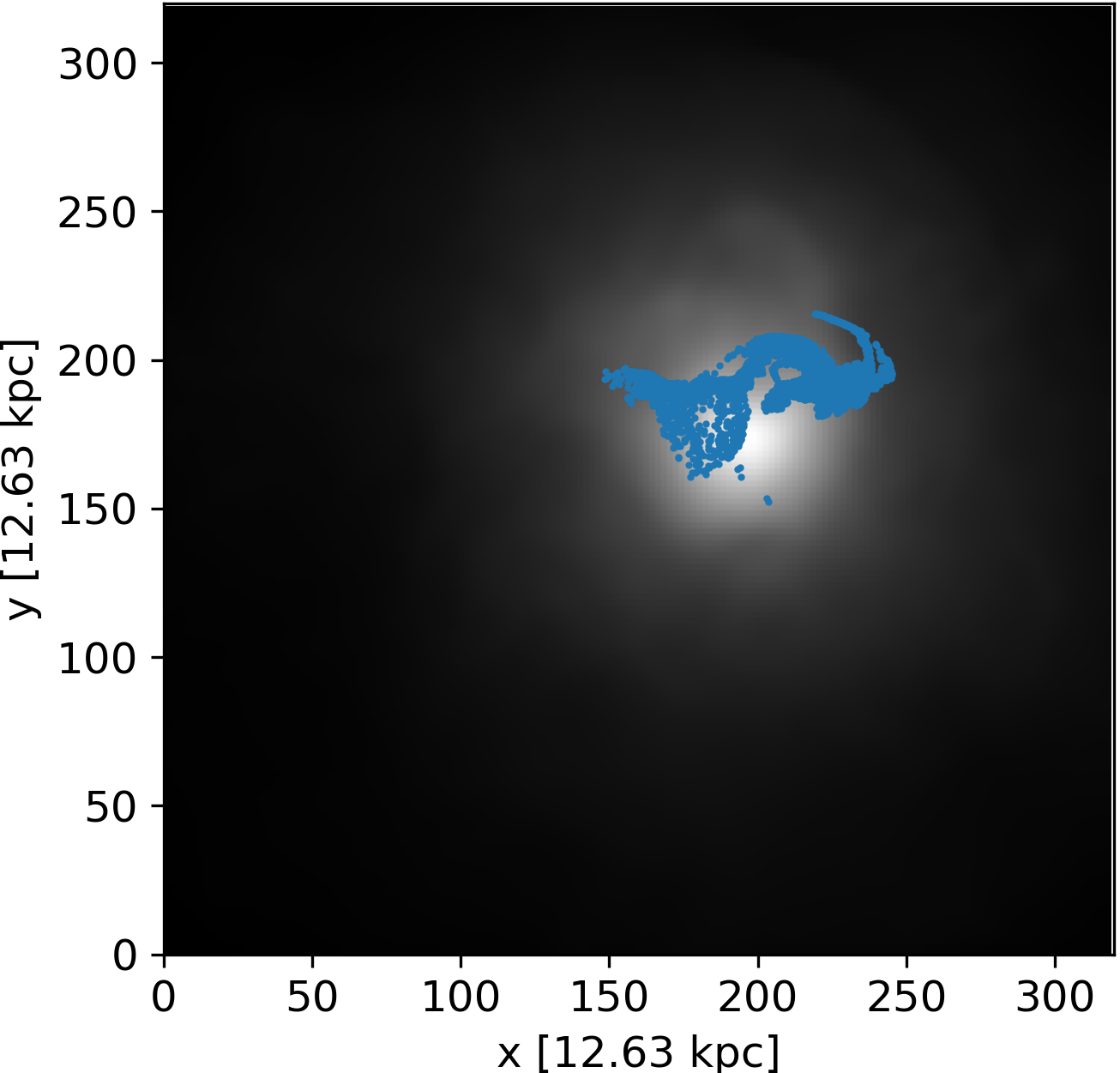}
    \label{subfig:evo_d}
    \caption{}
    \end{subfigure}

    \caption{Evolution of the projected density (in grey) overlaid with the projected position of the tracer particles forming the relic. The plot (a) at $z=0.119$ and (b) at $z=0.106$ show the formation of the relic. The plot (c) shows the moment of the highest luminosity of the relic at $z=0.071$. The plot (d) shows the evolution of the relic at $z=0$.}
    \label{fig:evo_of_relic}
\end{figure*}

To compare the radio luminosity and the acceleration efficiencies for single- and multishock scenarios, we searched for giant radio relics in the simulation using the HOP-algorithm (cf. Sec.~\ref{sec:HOP}). We have identified the largest and most powerful relic in our simulation. The relic forms at a time of $\sim \unit[12.8]{Gyr}$ (or redshift $z=0.071$) and has a length of $\unit[0.93]{Mpc}$. Moreover, the relic has a similar size and radio power as its observed counterparts. Hence, it is a good candidate for our study.
The relic is sampled by 4707 tracer particles. The spatial evolution and formation of the relic can be seen in Fig.~\ref{fig:evo_of_relic}.

\begin{figure}
    \centering
    \includegraphics[width=0.5\textwidth]{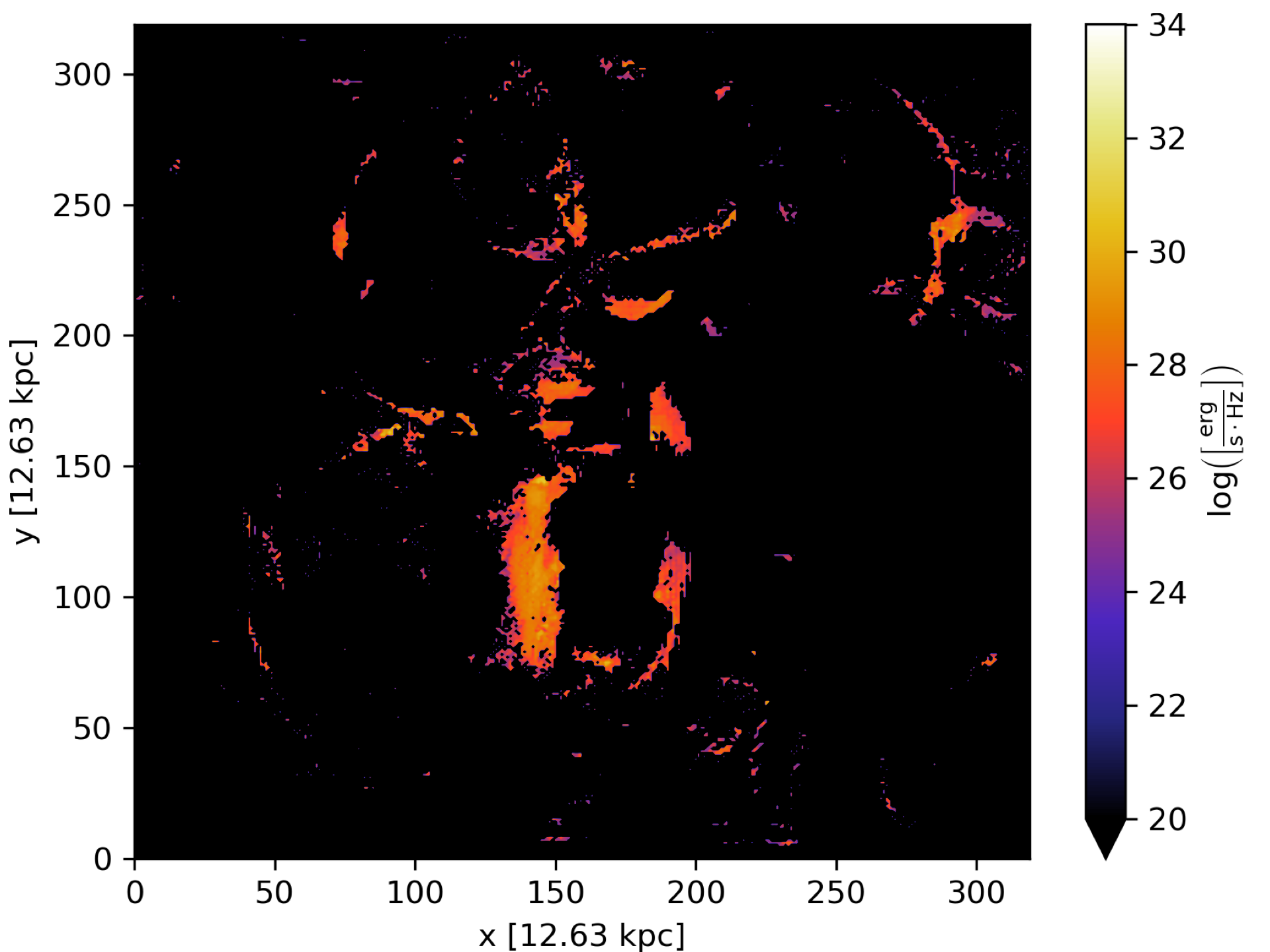}
    \caption{Radio emission of the entire galaxy cluster region, using the \citet{2007MNRAS.375...77H} model. The model is used to find structures inside the cluster by applying a HOP-Finder on the calculated radio emissions. The \citet{2007MNRAS.375...77H} radio emission is only used for finding structures. For the analysis of the relic, we used a more elaborated model, as described. The large structure near the centre corresponds to the relic we use for the analysis.}
    \label{fig:HB07_total_cluster}
\end{figure}
  We measured the magnetic field strength, the gas density, the Mach number and the gas temperature in the  relic's region, tracer-based. 
  The magnetic field strength in the radio relic is $\sim \unit[0.09\pm0.04]{\mu G}$. The thermal gas in this relic has temperatures around $\sim \unit[4.9\pm0.6\cdot 10^{7}]{K}$. The typical Mach number of the shocked tracers in the relic is $\sim 2.3$ with a standard deviation of $\sim 0.8$. The fraction of tracer particles in the relic that have experienced at least one shock in the $\unit[0.6]{Gyr}$ prior to the formation of the relic is around $89\%$.

In Fig.~\ref{fig:HB07_total_cluster}, we plot the radio emission of the whole cluster using the \citet{2007MNRAS.375...77H} model, which is used for finding the structures using the HOP halo finder. Since the axis scaling of Fig.~\ref{fig:evo_of_relic}~(c) and Fig.~\ref{fig:HB07_total_cluster} is the same, the selected relic can be identified with the structure around $x[\unit[12.63]{kpc}]\approx 150$ and between $y[\unit[12.63]{kpc}]\approx 75$ and $y[\unit[12.63]{kpc}]\approx 150$.
  
  \begin{figure*}
      \centering
      \begin{subfigure}{0.49\textwidth}
          \centering
          \includegraphics[width=1\linewidth, scale=1.0]{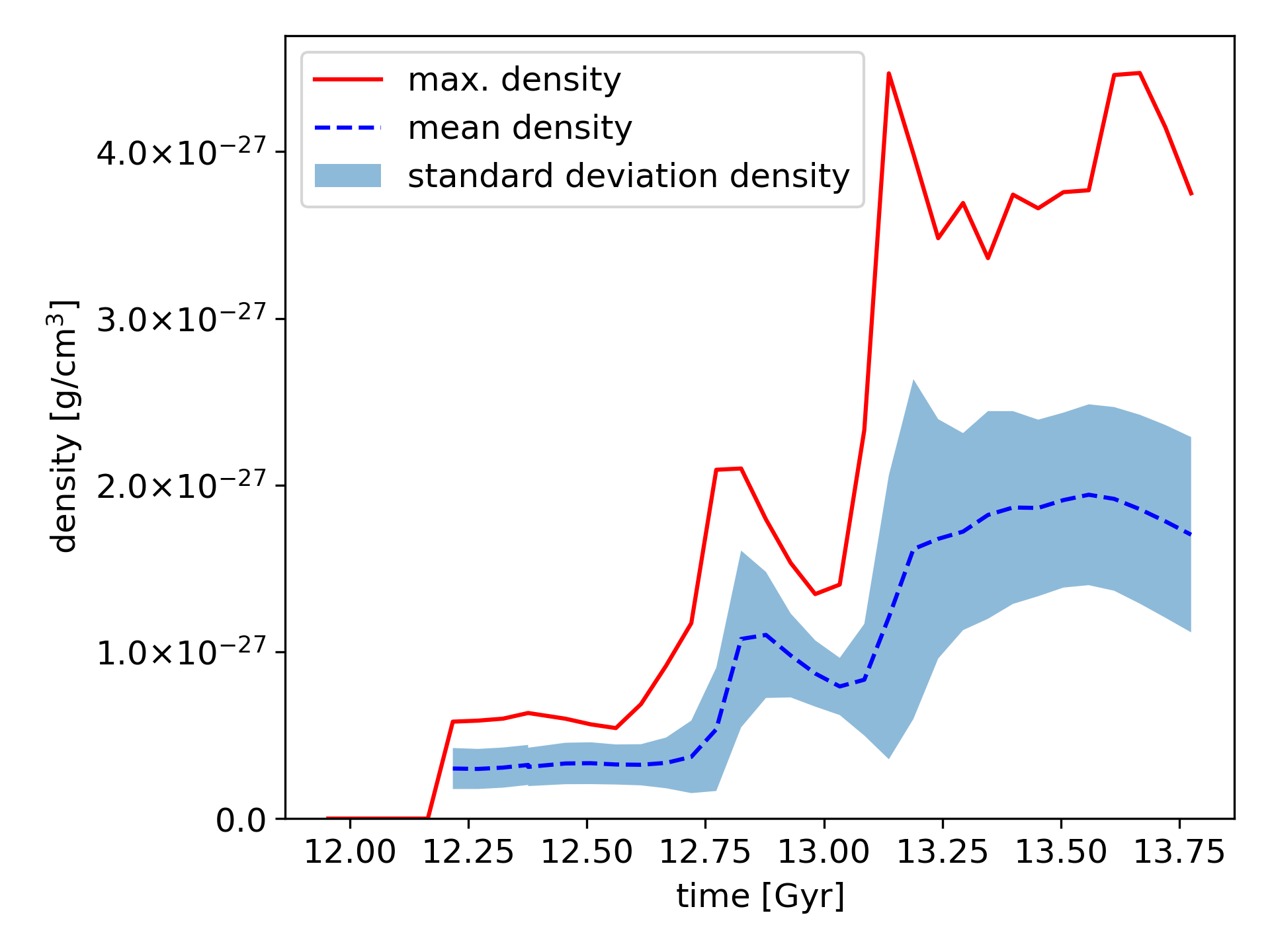}
          \label{fig:evo_quan_rho}
          \caption{}
      \end{subfigure}
      \begin{subfigure}{0.49\textwidth}
          \centering
          \includegraphics[width=1\linewidth, scale=1.0]{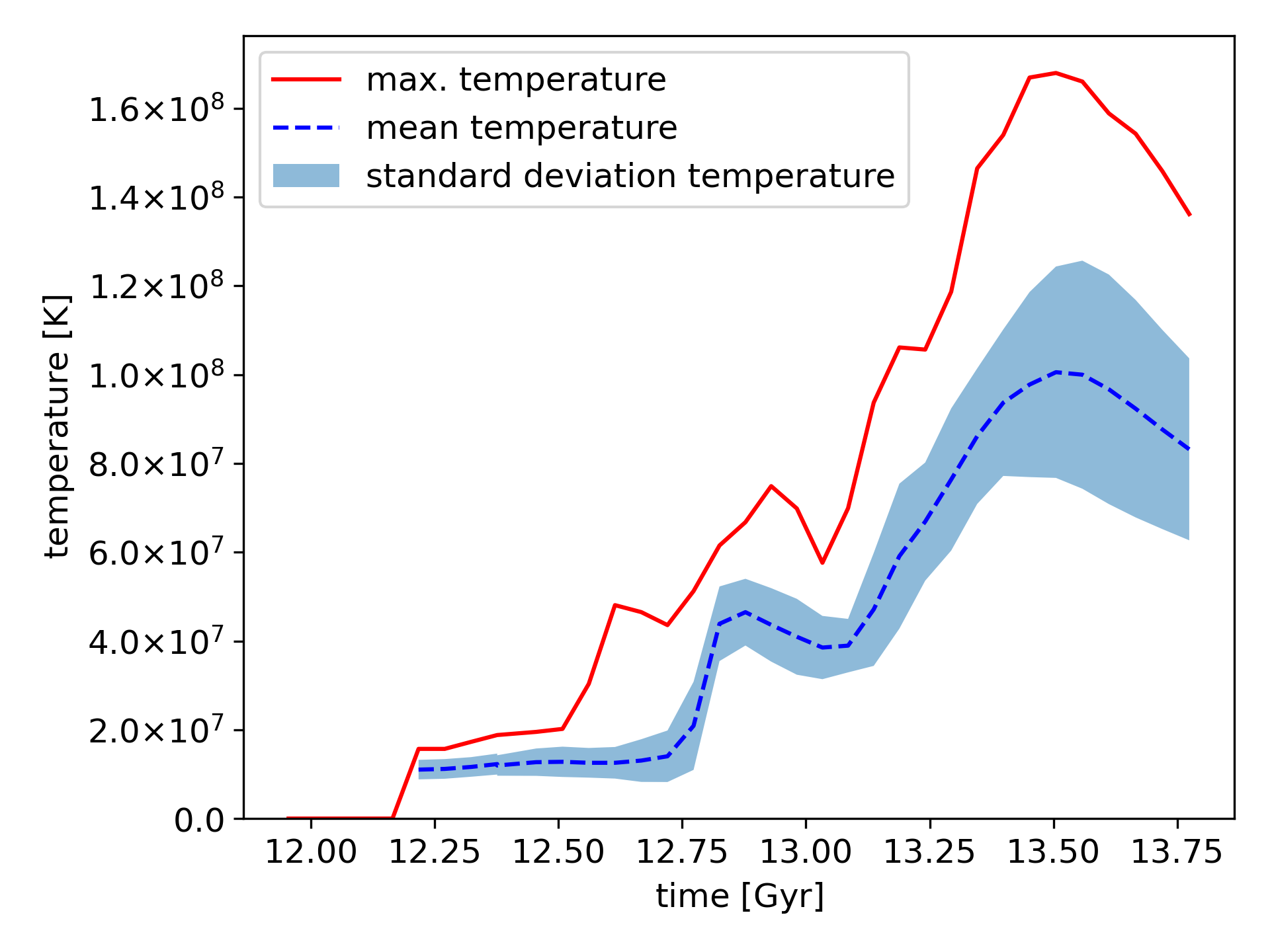}
          \label{fig:evo_quan_T}
          \caption{}
      \end{subfigure}
      \begin{subfigure}{0.49\textwidth}
          \centering
          \includegraphics[width=1\linewidth, scale=1.0]{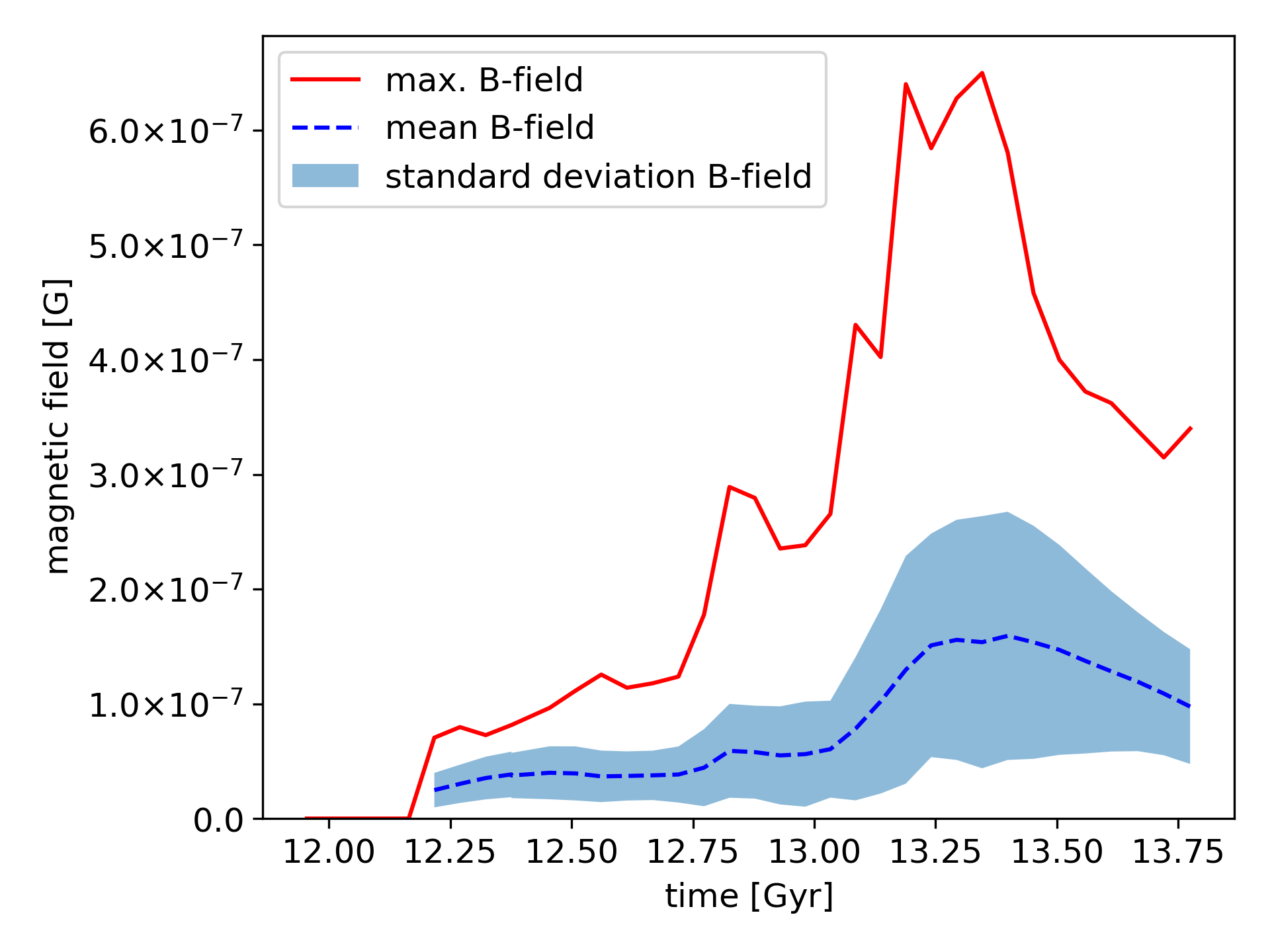}
          \label{fig:evo_quan_B}
          \caption{}
      \end{subfigure}
      \begin{subfigure}{0.49\textwidth}
          \centering
          \includegraphics[width=1\linewidth, scale=1.0]{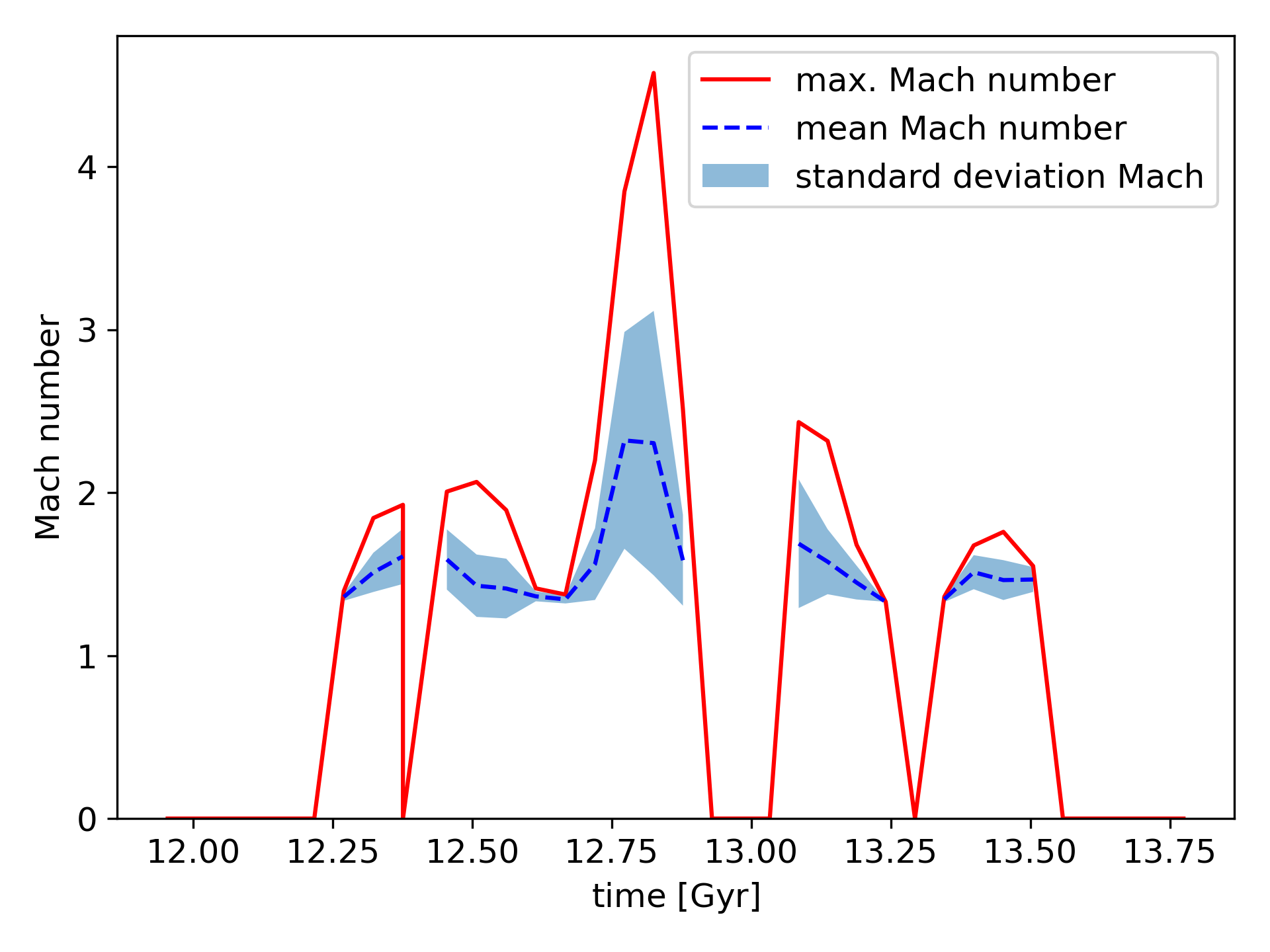}
          \label{fig:evo_quan_M}
          \caption{}
      \end{subfigure}
      \caption{Evolution of the mean value, the maximum value and the standard deviation for the density, temperature, magnetic field strength and Mach number (only the shocked tracers) for the tracers of the selected relic. The mean value is plotted as the blue line, the standard deviation is plotted as the blue area around the mean value, the maximum value is represented by the red line, for all quantities.}
      \label{fig:evolutionquantities}
  \end{figure*}
  
  \begin{figure}
    \centering
    \includegraphics[width=0.45\textwidth]{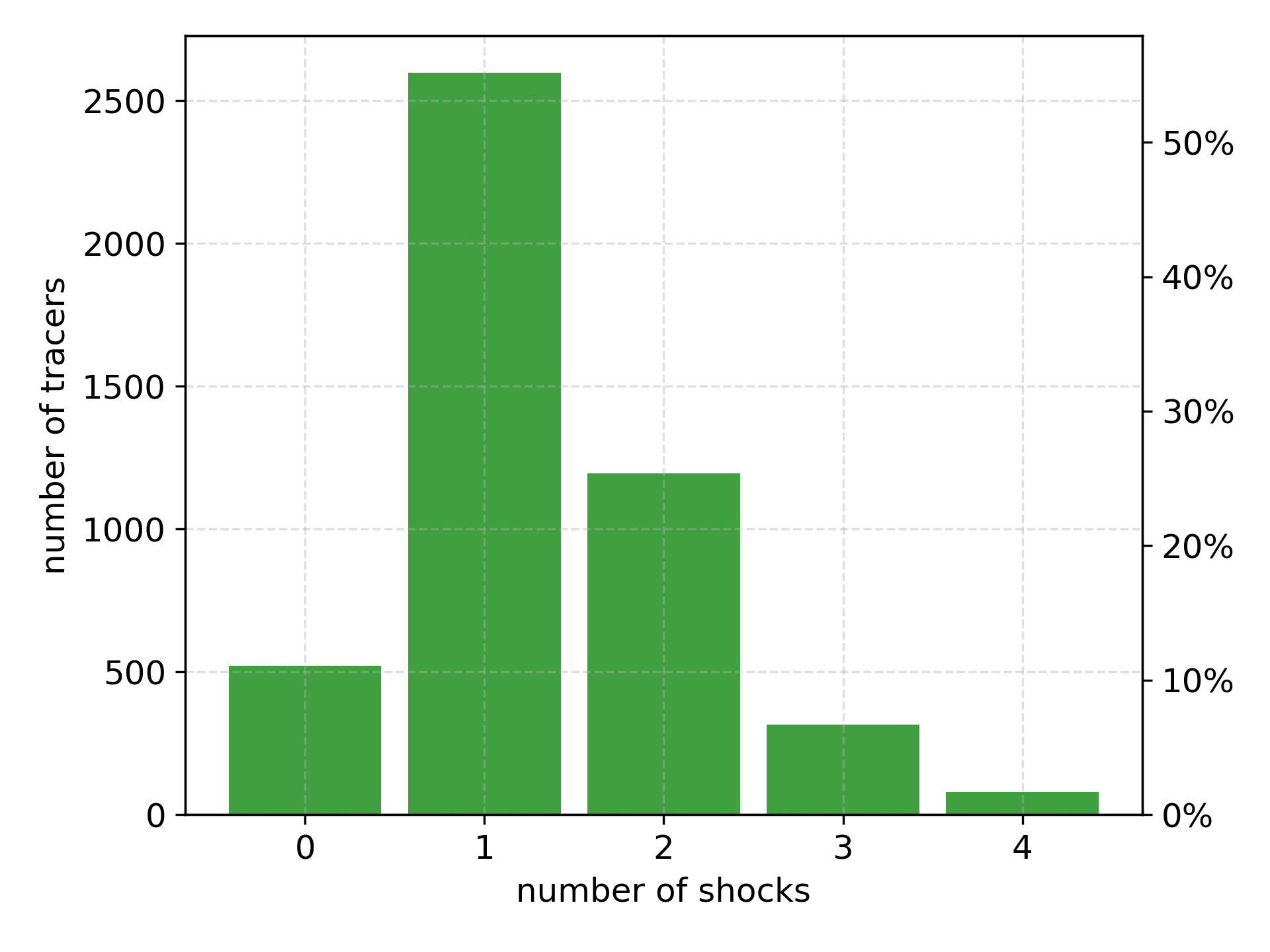}
    \caption{Histogram of the shock history for the particles forming the relic at $z=0.071$, the moment the relic appears.}
    \label{fig:shockhist}
\end{figure}
 
In Fig.~\ref{fig:evolutionquantities}, we plot the evolution of the density, the temperature, the magnetic field strength and Mach number of the tracers associated with the relic.
It is evident that the particles that make up the relic enter the high-resolution volume very late.  The first particles enter at $\unit[12.17]{Gyr}$. The Mach number of the shock peaks at $\unit[12.77]{Gyr}$, but particles also experience shocks between $\unit[12.17]{Gyr}$ and $\unit[12.77]{Gyr}$.  The passage of shocks can also be seen as peaks in the density and temperature plots.

In Fig.~\ref{fig:shockhist}, the histogram shows the number of times a tracer has experienced a shock in its lifetime (i.e. until $\unit[12.77]{Gyr}$). The majority ($\sim 55\ \%$) of particles have been shocked only once, i.e. by the shock that produces the relic itself. The second biggest fraction are particles that have experienced two shocks ($\sim 25\ \%$).
$\sim 11\ \%$ of particles have never experienced a shock but they still lie within the relic volume. $\sim 7\ \%$ have experienced three shocks, $\sim 2\ \%$  have experienced four shocks. No tracer particle has experienced more than four shocks. 

Following these results, the time range in which multiple shock events occur is below the cooling time of the CR electrons. Therefore it is plausible that the MSS has a significant impact on the evolution of the relic.

\subsection{The evolution of the electron spectrum}\label{sec:evo_e_spec}

Using the methods described in Sec.~\ref{sec:synchroemission}, we compute the evolution of the energy spectrum of the particles belonging to the relic.
We split the computation of the electron spectra into four cases: In the first case (model \textbf{A}) we used re-acceleration with no cut on the obliquity. In the second case (model \textbf{B}) we used re-acceleration in combination with a cut on the obliquity.  In the third case (model \textbf{C}) we used only acceleration from the thermal pool with no cut on the obliquity. In the fourth case (model \textbf{D}) we used only acceleration from the thermal pool in combination with a cut on the obliquity. In model \textbf{C} and \textbf{D}, we take only the last shock that a particle suffers into account. By only considering the last shock, the particles only experience direct acceleration and no re-acceleration. The last shock that a particle experiences coincides with the shock that is associated with the relic. For all models applies that there are multiple injections, owing to the continuous infall of matter. 
By using the obliquity cut according to Eq.~\ref{eq:obliquity_cut}, only quasi-perpendicular shocks are considered. In the case that no obliquity cut is used, the dependence on $\Theta$ for acceleration efficiency is omitted, and Eq.~\ref{eq:obliquity_cut} simplifies to $\eta(M,\Theta)=\tilde{\eta}(M)$.
\begin{figure}
    \centering
    \includegraphics[width=0.45\textwidth]{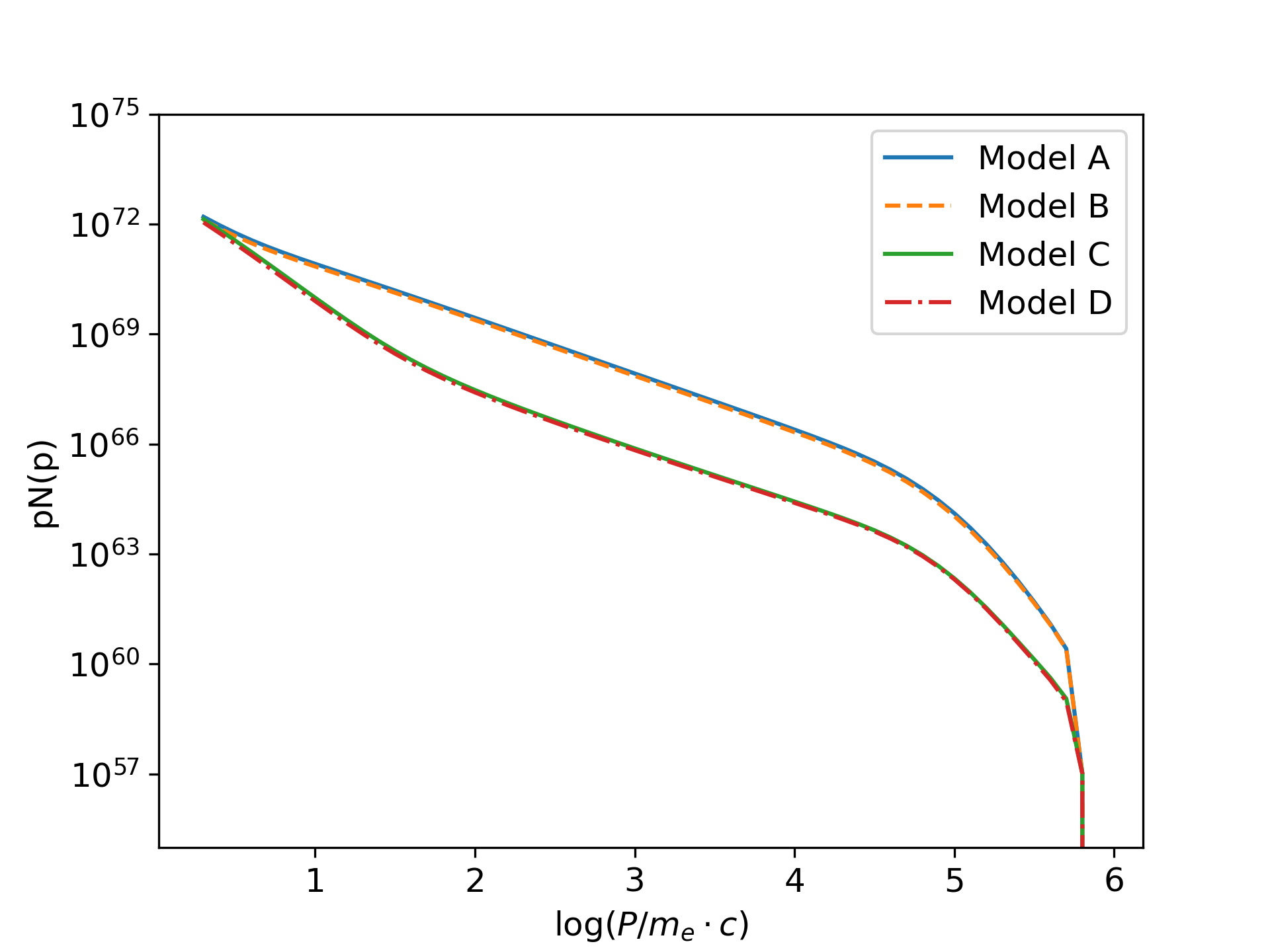}
    \caption{Electron spectra for the tracers composing the relic at redshift $z=0.071$. The blue line (model A) represents the case of a relic using re-acceleration without an obliquity cut, the orange (model B) line represents the case of a relic using re-acceleration in combination with an obliquity cut, the green line (model C) represents the case of a relic using acceleration without an obliquity cut, the red line (model D) represents the case of a relic using acceleration in combination with an obliquity cut.
    }
    \label{fig:electronspec}
\end{figure}
Fig.~\ref{fig:electronspec} shows the electron spectrum at $z=0.071$, the time at which the relic appears. For all four spectra, we find a spectral index in the range of $\delta_\mathrm{inj}=2.8\pm0.6$. This corresponds to the Mach numbers for radio relic, which should be between $\sim2-3$.
The difference between the acceleration spectra and the spectra including re-acceleration is substantial. We see in the middle section of the plot a difference between the spectra as big as $10^2$. Regardless of whether re-acceleration is used or not, the spectrum is a factor of three smaller when the obliquity cut is used.

\subsection{Radio luminosity of the relic}\label{sec:radio_emission_result}

\begin{figure}
    \centering
    \includegraphics[width=0.5\textwidth]{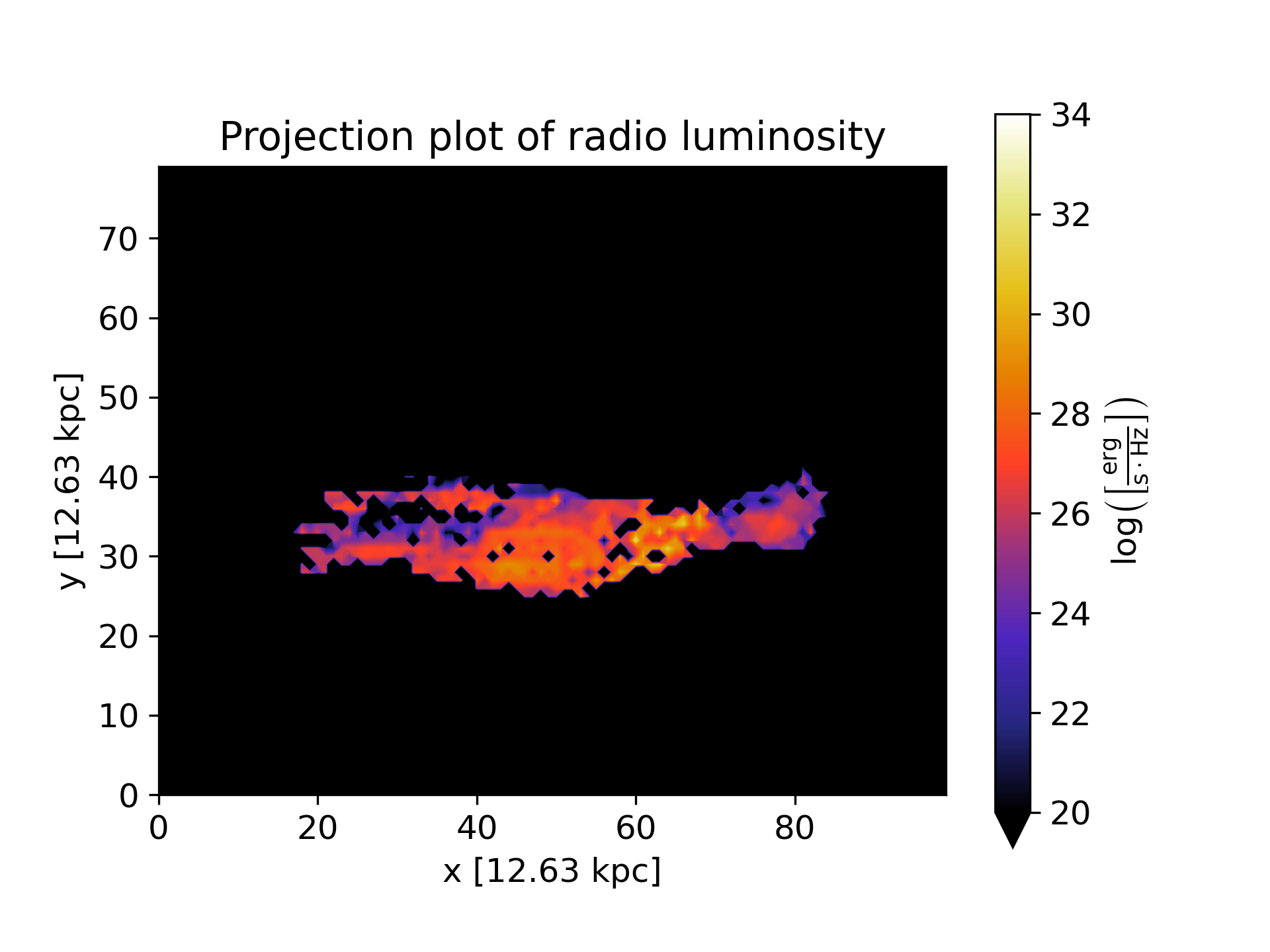}
    \caption{Radio emission of the relic for the case of re-acceleration with an obliquity cut (model B). For a better view, the plot is turned by 90 degrees in comparison to Fig.~\ref{fig:evo_of_relic}.}
    \label{fig:radio_all_shock}
\end{figure}

In this section, we calculated the synchrotron luminosity for each tracer at a frequency of $\unit[1400]{MHz}$.
Fig.~\ref{fig:radio_all_shock}, shows a map of the relic at $\unit[1400]{MHz}$.
As described in Sec.~\ref{sec:evo_e_spec} we examine four different cases. For these cases, the total radio luminosities are listed in Tab.~\ref{tab:radiolumshock}. In addition, we split the contribution to the total radio luminosity of the tracers by the number of shocks experienced for the four different cases. This can be seen in Tab.~\ref{tab:radiolumshock}.

\begin{table}
    \centering
    \caption{Radio luminosity for the different cases split up by the number of shocks experienced. In the case of acceleration a tracer particle experiences only the last shock, there are no tracers that experience more than one shock.}
    \label{tab:radiolumshock}
    \begin{tabular}{c|c|c|c|c}
         \hline
         num. of shocks & Model A & Model B & Model C & Model D\\
         & $\bigl[\mathrm{\frac{erg}{s\cdot Hz}}\bigr]$ & $\bigl[\mathrm{\frac{erg}{s\cdot Hz}}\bigr]$ & $\bigl[\mathrm{\frac{erg}{s\cdot Hz}}\bigr]$ & $\bigl[\mathrm{\frac{erg}{s\cdot Hz}}\bigr]$\\
         \hline
         1 &  $3.2\cdot10^{30}$ &  $2.8\cdot10^{30}$ &  $3.5\cdot10^{30}$ &  $3.0\cdot10^{30}$\\
         2 & $2.7\cdot10^{31}$ & $1.5\cdot10^{31}$ & - & - \\
         3 & $1.1\cdot10^{32}$ & $9.3\cdot10^{31}$ & - & - \\
         4 & $3.7\cdot10^{31}$ & $3.7\cdot10^{31}$ & - & -\\
         \hline
         total luminosity & $1.8\cdot10^{32}$ & $1.5\cdot10^{32}$ & $3.5\cdot10^{30}$ & $3.0\cdot10^{30}$\\
         \hline
    \end{tabular}
\end{table}

In models \textbf{A} and \textbf{B}, the tracers that have experienced three shocks produce the highest radio luminosity. In both cases, the luminosity of the particles shocked three times makes up about $\sim 62\%$ of the total luminosity of the relic. 
However, these particles constitute only a small fraction, $\sim 8 \%$, of the particles that contribute to the radio emission of the relic (not including the particle that have not been shocked).
The obliquity also has an impact on the radio luminosity because fewer particles experience acceleration and re-acceleration.
In the acceleration cases (\textbf{C} and \textbf{D}), the total luminosity of the relic has just $2\%$ of the luminosity of the relic with re-acceleration. 
However, in the case of acceleration (model \textbf{C} and \textbf{D}), the luminosity of particles shocked once is bigger than in the re-acceleration case since in the simple acceleration case all particles are just shocked once. We have also made spatial comparisons of the radio luminosity to analyze which part of the radio luminosity belongs to which tracer family.
Compared to the ratios between the re-acceleration and acceleration models, the ratios between the cases with and without obliquity cut are rather small and in the range of  $\sim 85\ \%$.

\begin{figure*}
    \centering
    \begin{subfigure}{0.45\textwidth}
        \centering
        \includegraphics[width=0.95\linewidth, scale=1.0]{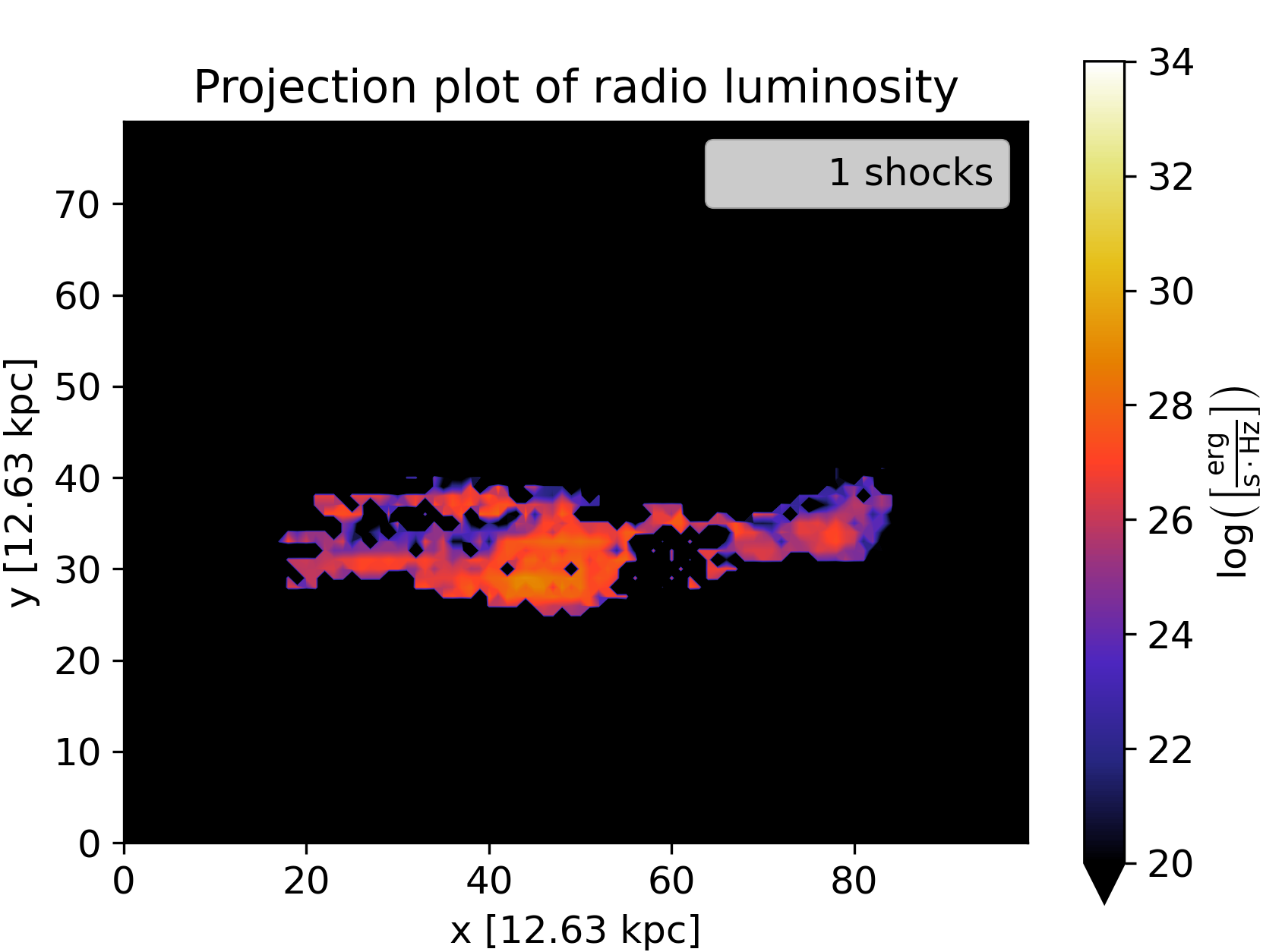}
        \caption{}
        \label{subfig:1shockradiolum}
    \end{subfigure}
    \begin{subfigure}{0.45\textwidth}
        \centering
        \includegraphics[width=0.95\linewidth, scale=1.0]{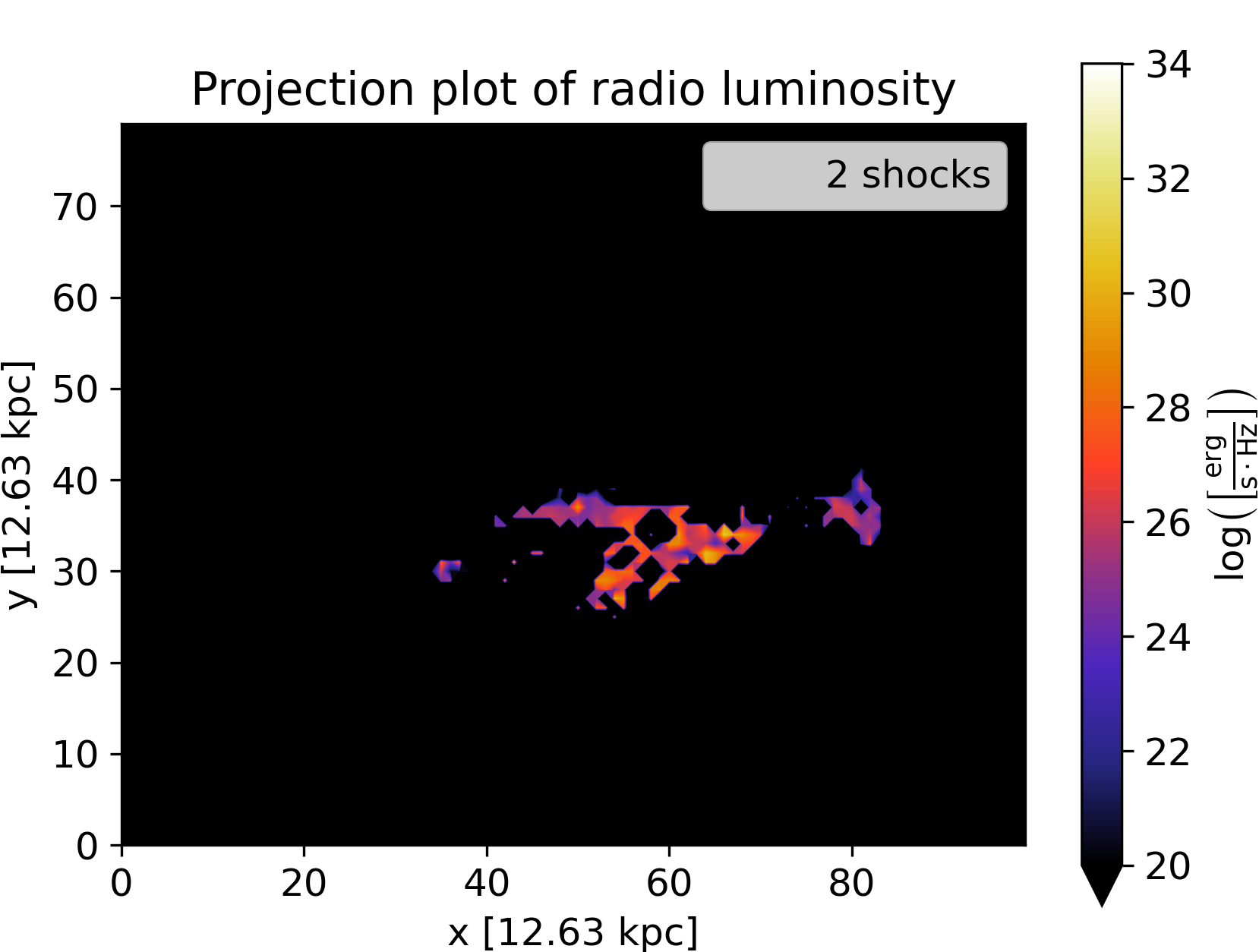}
        \caption{}
        \label{subfig:2shockradiolum}
    \end{subfigure}
    \begin{subfigure}{0.45\textwidth}
        \centering
        \includegraphics[width=0.95\linewidth, scale=1.0]{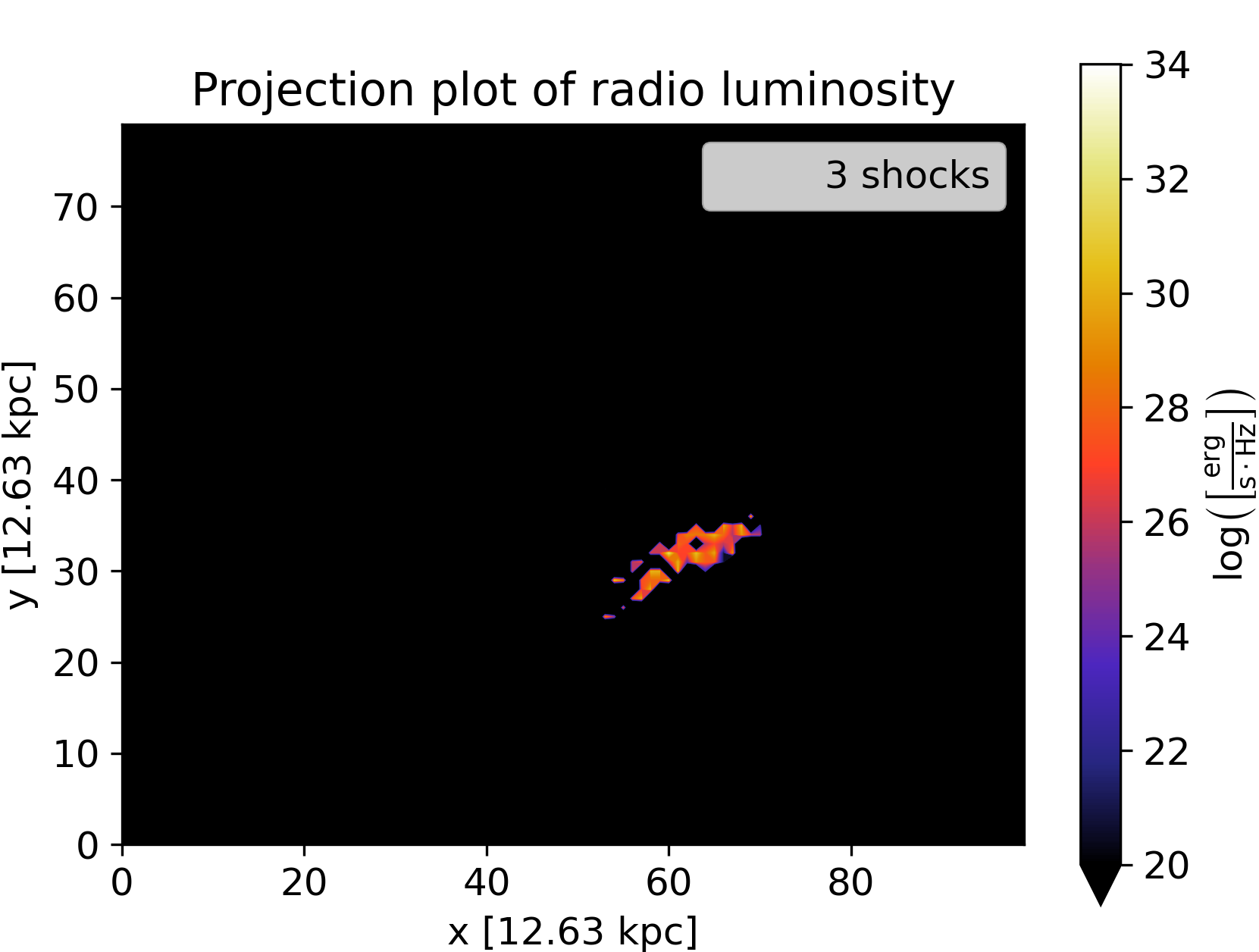}
        \caption{}
        \label{subfig:3shockradiolum}
    \end{subfigure}
    \begin{subfigure}{0.45\textwidth}
        \centering
        \includegraphics[width=0.95\linewidth, scale=1.0]{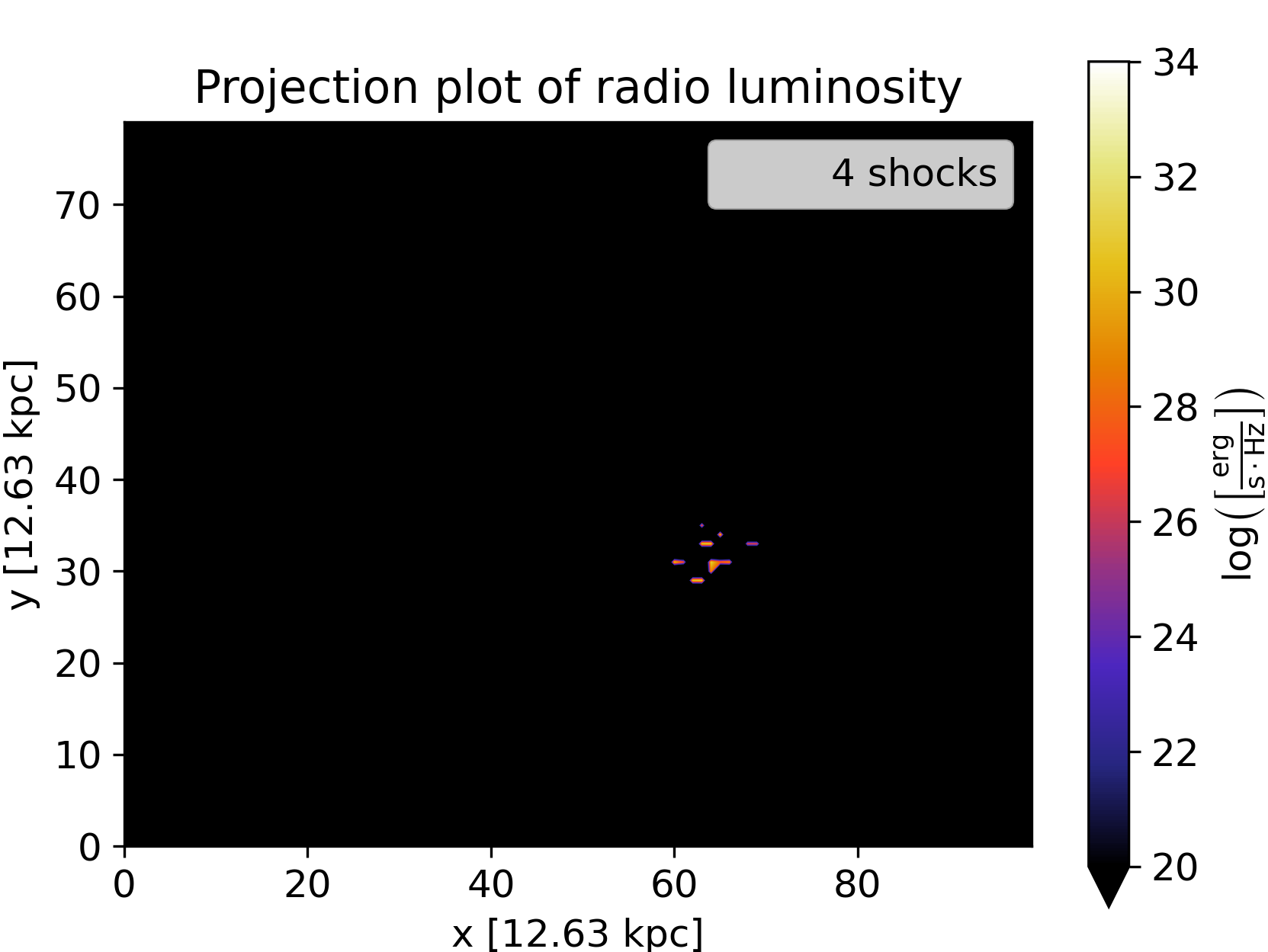}
        \caption{}
        \label{subfig:4shockradiolum}
    \end{subfigure}
    \caption{Showing the radio luminosity of the relic filtered for the tracer families. Panel (a) shows the radio luminosity caused by tracers that are shocked once, (b) twice, (c) three times and (d) four times. The tracers that have suffered three shocks contribute most to the radio luminosity. Again, the plot is turned by 90 degrees in comparison to Fig.~\ref{fig:evo_of_relic}.}
    \label{fig:radiolumbyshock}
\end{figure*}
By comparing Fig.~\ref{fig:radio_all_shock} and \ref{fig:radiolumbyshock}, we can evaluate where each tracer family 
is located within the relic.
Both, the family of thrice shocked tracers and the family of tracers that have been shocked four times are locally confined.
On the other hand, the family of the tracers shocked only once is spread across the entire relic region.
The family of twice-shocked tracers occupies a small subregion inside the relic.

\begin{figure}
    \centering
    \includegraphics[width=0.5\textwidth]{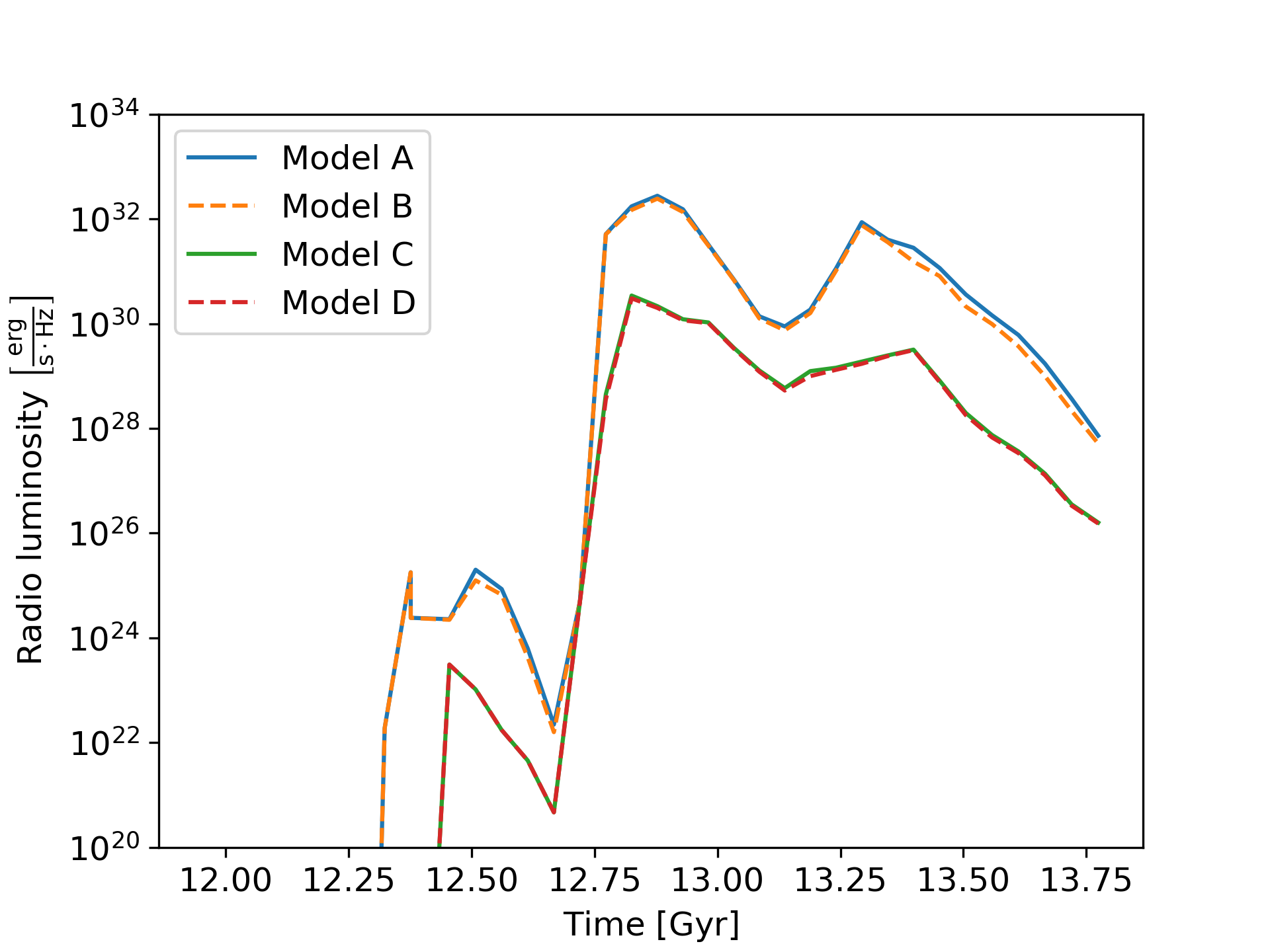}
    \caption{Evolution of the radio luminosity for models A-D. The relic appears at t=$\unit[12.77]{Gyr}$ as seen in the steep rise of the luminosity.} 
    \label{fig:time_evo_radio}
\end{figure}

We also computed the evolution of the radio luminosity of the relic, see Fig.~\ref{fig:time_evo_radio}.
In the evolution, the luminosity of the two models that use only acceleration (\textbf{C} and \textbf{D}) is lower than in the models that also use re-acceleration (\textbf{A} and \textbf{B}).
The importance of re-acceleration can be seen in the offset between the lines at the time when the luminosity increases strongly at $t=\unit[12.77]{Gyr}$. 
Models with re-acceleration are by a factor of $\sim 10^2-10^{3.5}$ more luminous than those without re-acceleration.
On the other hand, the difference between the cases with and without obliquity cut is small over time. The difference between the models are in the range of $\sim 1-3$.

\subsection{Acceleration efficiencies}\label{sec:acc_eff_result}

We now compare the input acceleration efficiencies to those that would be inferred from the resulting radio luminosities.

In order to calculate the acceleration efficiencies and compare the measured efficiencies to the input efficiencies of our model, we follow the approach of \citet{2020A&A...634A..64B}. 
The acceleration efficiencies measures the fraction of kinetic energy dissipated at the shock that goes in the acceleration of CR. This is shown by equation \ref{eq:energyatshock}.
Re-arranging Eq. \ref{eq:HB07_without_prefactors} gives us an expression for the acceleration efficiency as in \citet{2020A&A...634A..64B}:

\begin{equation}\label{eq:HB07}
\begin{split}
    \eta_\mathrm{e}(M)=\frac{P_{1.4}}{C\,A n_\mathrm{e,-4}\frac{\xi_\mathrm{e}}{0.05}\left(\frac{T_\mathrm{d}}{\unit[7]{keV}}\right)^{\frac{3}{2}}\nu_\mathrm{1.4}^{-\frac{s}{2}} }\cdot \frac{\left(\frac{B_\mathrm{CMB}}{\unit[]{\mu G}}\right)^2+\left(\frac{B}{\unit[]{\mu G}}\right)^2}{\left(\frac{B}{\mu G}\right)^{1+\frac{s}{2}}} .
\end{split}
\end{equation}
In this formula, $P_{1.4}$ represents the radio power at 1.4 GHz that we obtained from the calculation of the synchrotron radiation. $n_\mathrm{e,-4}$ is the electron number density in units of $10^{-4}$ cm$^{-3}$. All other variables are the same as we introduced them in Sec.~\ref{sec:synchroemission}. In the form presented here, $C=\unit[6.4\cdot10^{34}]{erg/(s \cdot Hz)}$. This constant is valid under the assumption of the further taken normalizations on $A$, $n_\mathrm{e}$, $\xi_\mathrm{e}$, $T_\mathrm{d}$ and both magnetic fields.
Using this approach, we obtain the acceleration efficiency determined directly from the radio luminosity, as it is done for real observations.\\

In our particle-based approach, there is a distribution of efficiencies because each tracer particle has its own acceleration efficiency. Comparing the distribution of efficiencies to the efficiency determined for the entire relic from the total luminosity is not very meaningful. Therefore, the spectral index was determined for the relic, from which the radio Mach number can be determined, as $M_\mathrm{radio}=\sqrt{(1-s)/(-1-s)}$. For this Mach number, we derive the acceleration efficiency of the underlying model. For our simulation, we model the efficiencies as in \citet{2013ApJ...764...95K} so that we can then compare the input and output efficiencies.

In equation \ref{eq:HB07}, we used the radio-weighted average values of the magnetic field and temperature, measured by the tracers.
Here, we used the radio-weighted averages, because they are biased to the brighter parts of the relics and, hence, better characterise what would be first picked up by observations.
We computed the radio spectral index using the seven frequencies $\nu= \unit[50, 140, 650, 1400, 5000, 10000]{MHz}$. 
In the following, we will mark this efficiency with $\eta_{e}^\mathrm{obs}$.

\begin{table*}
    \centering
    \caption{Showing the acceleration efficiencies with the underlying quantities for the four models. In the upper two sections the acceleration efficiencies are calculated from the radio luminosity, following the approach that is also used in observations. In the first section, we used arithmetic mean values for the computation of the efficiency. In the section below, we used the radio-weighted average. In the third section, we computed the efficiencies that went into the model. In the last section, we compared the measured efficiencies to the efficiencies the model used.}
    \label{tab:acc_efficiencies_all}
    \begin{tabular}{c|c|c|c|c}
        \hline
         & Model A & Model B & Model C & Model D  \\
         & reacc. no oc. & reacc. oc. & acc. no oc. & acc. oc.\\
         \hline
         radio-weighted average\\
         \hline
         $\langle T_d  \rangle$ & $\unit[4.9\cdot10^7]{K}$ &  $\unit[4.9\cdot10^7]{K}$ & 
         $\unit[4.1\cdot 10^7]{K}$ & $\unit[4.2\cdot10^7]{K}$\\
         $\langle B \rangle$ & $\unit[8.6\cdot10^{-2}]{\mu G}$ & $\unit[8.6\cdot10^{-2}]{\mu G}$ & $\unit[1.3\cdot10^{-1}]{\mu G}$ &
         $\unit[1.3\cdot10^{-1}]{\mu G}$ \\
         $\eta_e^\mathrm{obs}$ & $7.2\pm0.2$ & $6.7\pm0.2$ & $(6.7\pm 0.1)\cdot10^{-2}$ & $(5.7\pm0.1)\cdot10^{-2}$\\
         \hline
         model acceleration efficiencies\\
         \hline
         $s_\mathrm{radio}$ & $1.40\pm0.02$ & $1.41\pm0.02$ & $1.25\pm0.02$ & $1.25\pm0.02$\\
         Mach & $2.5\pm0.1$ & $2.4\pm0.1$ & $3.0\pm0.1$ & $3.0\pm0.1$\\
         $\eta_e^\mathrm{model}$ & $(3.7\pm0.5)\cdot 10^{-3}$ & $(3.4\pm0.5)\cdot 10^{-3}$ & $(8.0\pm1.0)\cdot 10^{-3}$ & $(8.0\pm1.0)\cdot 10^{-3}$\\
         \hline
         ratio of model to obs. efficiencies\\
         \hline
         $\eta_e^\mathrm{obs}/\eta_e^\mathrm{model}$ & $1945.9$ & $1970.6$ & $8.4$ & $7.1$\\
         \hline
    \end{tabular}
\end{table*}

We summarized the inferred acceleration efficiencies for the different models in Tab.~\ref{tab:acc_efficiencies_all}. For the models that include re-acceleration, the inferred efficiencies are too high for a standard DSA scenario and they would violate energy constraints.

In model \textbf{A}, the acceleration efficiency is $\eta_e^\mathrm{obs}=7.17\pm0.21$, whereas in model \textbf{B}, with the cut on the obliquity, the acceleration efficiency is $\eta_e^\mathrm{obs}=6.71\pm0.20$.\\
Observations show that acceleration efficiencies greater than one can be found for many of the known radio relics \citep[e.g.][]{2020A&A...634A..64B}.
On the other hand, the acceleration efficiencies for the models using only acceleration (\textbf{C} and \textbf{D}) are in the range of $\sim (5.6-6.8)\cdot 10^{-2}$, so they are physical and reasonable for DSA from the thermal pool.
The acceleration efficiency values  following the approach of \citet{2020A&A...634A..64B} ($\eta_e^\mathrm{obs}$ derived from Eq.~\ref{eq:HB07}) of model \textbf{A} and \textbf{B} are clearly unphysically high. In the re-acceleration model, the Mach number and acceleration efficiency of the shock cannot be inferred from the radio spectral index and the radio power, respectively, based on the expectation of the simple version of DSA model. This is because the CRe do not come from the dissipated kinetic energy at the shock, cf. Eq.~\ref{eq:energyatshock}. This means that the right-hand side of Eq.~\ref{eq:energyatshock} changes because $E_\mathrm{CR}$ is much larger since there is an additional power-law tail in the integral of $E_\mathrm{CR}$ that comes from a previous acceleration. Equating this boosted $E_\mathrm{CR}$ to the left-hand side of Eq.~\ref{eq:energyatshock}, the left-hand side, i.e., the acceleration efficiency $\eta$ then becomes quite high.

By comparing the efficiencies of model \textbf{A} to model \textbf{C} and model \textbf{B} to model \textbf{D}, we see how re-acceleration changes the observed acceleration efficiency. In the radio-weighted case the ratio between the acceleration efficiency of model \textbf{A} to model \textbf{C} is $\eta_e^\mathrm{obs, A}/\eta_e^{\rm obs, C}\approx107$, the ratio between model \textbf{B} to model \textbf{D} is $\eta_e^\mathrm{obs, A}/\eta_e^{\rm obs, C}\approx118$. 
In contrast to those values, the obliquity cut itself does not show big differences in the efficiencies and therefore the ratio between model 
\textbf{A} and \textbf{B} and also between model \textbf{C} and \textbf{D} is small.

Moreover, we have the knowledge of which acceleration efficiencies have been included in the simulation. We can compute the acceleration efficiency directly by calculating the radio Mach number from the spectral index and inserting it into the model of \citet{2013ApJ...764...95K}. 
In the following, we will call this efficiencies $\eta_e^\mathrm{model}$. The determination of $\eta_e^\mathrm{model}$ allows us to easily compare the acceleration efficiencies that went into the model to the acceleration efficiencies one would measure, i.e. $\eta_e^\mathrm{obs}$.

In the third section of Tab.~\ref{tab:acc_efficiencies_all}, we present the efficiencies that went into the simulation as described previously.
For the models with re-acceleration, the acceleration efficiencies that went into the simulation are $\eta_e^\mathrm{model}=(3.7\pm0.5)\cdot10^{-3}$ and $\eta_e^\mathrm{model}=(3.4\pm0.5)\cdot10^{-3}$ for model \textbf{A} and \textbf{B}, respectively. For the models only using acceleration, the values are slightly higher, $\eta_e^\mathrm{model}=(8.0\pm1.0)\cdot10^{-3}$ for, both, model \textbf{C} and \textbf{D} since the radio Mach number inferred from the spectral index is higher.\\

We now compare the efficiencies one would observe to those of the underlying model. We calculated the ratios, the results are shown in the fourth section in Tab.~\ref{tab:acc_efficiencies_all}.
In model \textbf{A}, the observed acceleration efficiency is $\sim1946$ times higher than the acceleration efficiency that went into the model. In model \textbf{B}, the observed acceleration efficiency is around $\sim1971$ times higher than the model acceleration efficiency. In contrast, such high ratios do not show up in the acceleration models \textbf{C} and \textbf{D}. The ratios here reach maximal values of $8.4$. The obliquity cut again plays a minor role.
In general, for brighter radio relics, a typical observed radio power is $\sim \unit[10^{32}]{erg/(s\cdot Hz)}$. A radio power of $\sim \unit[10^{30}]{erg/(s\cdot Hz)}$, on the other hand, is very low and unlike most of known radio relics. Therefore, it can be assumed that also for real radio relics the MSS plays a major role. The ratios of acceleration efficiencies suggest that for many real radio relics, the actual acceleration efficiencies may be below the (large) acceleration efficiencies required to match the observed emission with the DSA model, under the simplistic assumption of direct acceleration by weak shock waves.

\section{Conclusions}\label{sec:conclusion}

In this paper, we studied the evolution of radio relics in a cosmological MHD simulation. Our aim was to test the influence of multiple shocks on the luminosity and the observed acceleration efficiencies in radio relics. We used Lagrangian tracer particles to follow the evolution of shock-accelerated CR electrons in the simulation. Applying a novel HOP halo finder, we selected all tracers that produce a radio relic. The simulated relic has properties that are similar to its observed counterparts. We used a Fokker-Planck solver to follow the evolution of the electron spectra under the influence of cooling and re-acceleration. Using the corresponding synchrotron emission, we investigated the underlying acceleration efficiencies. 

For many relics, the estimated acceleration efficiencies are unphysically high \citep[e.g.][]{2020A&A...634A..64B}. 
In extreme cases, these relics show acceleration efficiencies larger than one, implying that energy conservation is violated. 

We only focus on the re-acceleration of fossil electrons in a MSS. The diffusion of CR electrons is not included. The combination of a cosmological simulation and Lagrangian tracer particles allowed us to follow the evolution of shock (re-)accelerated cosmic rays during a galaxy cluster merger. 
 In MSS, the relic luminosity is $\sim 50$ times larger than in the case of acceleration from the thermal pool. Hence, we confirm the results by \citet{2022MNRAS.509.1160I}. However, in our case, the relic emission is dominated by particles that have experienced three shocks and not two shocks, as seen by \citet{2022MNRAS.509.1160I}.  
The three-shock family of tracers makes up $\sim 62 \ \%$ of the luminosity of the relic. 
Confining acceleration and re-acceleration to quasi-perpendicular shocks did not significantly affect the relic's luminosity. Hence, the obliquity seems to play a minor role in the MSS.\\

In the second step of our analysis, we compared the acceleration efficiencies inferred from the radio luminosity (i.e. the mock-observed efficiency, cf. Eq.~\ref{eq:HB07}) to the acceleration efficiencies from the underlying model. If the relic forms in a MSS, the observed acceleration efficiency is $7.2\pm0.2$. This apparent efficiency is $1.9\cdot10^3$ times larger than the efficiency of the underlying model. 
On the other hand, if the relic forms in a single-shock scenario, i.e., particles experience only one shock, the observed efficiency is $(6.7\pm0.1)\cdot10^{-2}$. The ratio between observed efficiencies and the model efficiencies are in our case smaller than $8.5$.
Again, the obliquity has little influence on the results.\\

We conclude that if a relic is dominated by re-accelerated electrons, the radio luminosity is significantly boosted and the inferred acceleration efficiency is larger than the actual efficiency of the shock acceleration process. The reason behind the high acceleration efficiencies lies in the interpretation of the (mock) observations ($\eta_\mathrm{obs}$). For calculating the acceleration efficiencies from observations, it is assumed that only direct acceleration plays a role. If re-acceleration is taken into account, the acceleration efficiency drops significantly ($\eta_\mathrm{model}$). On the other hand, if a relic is produced by acceleration from the thermal pool, the inferred acceleration efficiency mirrors the actual efficiency of the shock acceleration. Hence, the determination of the acceleration efficiency is a non-trivial task. Especially in the case of re-acceleration, $\eta$ cannot be derived from observations in the customary manner because we do not know the acceleration history of the cosmic rays. We note that this should be independent of the origin of re-accelerated particles', i.e. previous shock acceleration or ejection from AGN.

\section*{Acknowledgements}

We thank the referee for a very constructive report.
DCS acknowledges support from the team of the Hummel-Cluster at the Regionales Rechenzentrum of the University of Hamburg for providing the computing infrastructure and giving helpful feedback. 
DW is funded by the Deutsche Forschungsgemeinschaft (DFG,
German Research Foundation) - 441694982.
MB acknowledges support from the Deutsche Forschungsgemeinschaft under Germany's Excellence Strategy - EXC 2121 "Quantum Universe" - 390833306.
FV acknowledges the financial support by the H2020 initiative, through the ERC StG MAGCOW (n. 714196) and from the Cariplo "BREAKTHRU" funds Rif: 2022-2088 CUP J33C22004310003.
The authors gratefully acknowledge the Gauss Centre for Supercomputing e.V. (www.gauss-centre.eu) for supporting this project by providing computing time through the John von Neumann Institute for Computing (NIC) on the GCS Supercomputer JUWELS at Jülich Supercomputing Centre (JSC), under project no. hhh44 (PI Denis Wittor).
Finally we want to acknowledge the developers of the following python packages, which were used extensively during this project: \texttt{NUMPY} \citep{2020Natur.585..357H}, \texttt{SCIPY} \citep{2020SciPy-NMeth}, \texttt{ASTROPY} \citep{2022ApJ...935..167A}, \texttt{YT\char`_ASTRO\char`_ANALYSIS} \citep{yt.astro.analysis, yt}, \texttt{MATPLOTLIB} \citep{Hunter:2007} and \texttt{H5PY} \citep{Collete2013-tm}.

\section*{Data Availability}
The data underlying this article will be shared on reasonable request to the corresponding author.
The ROGER code used to evolve the momentum spectra of relativistic electrons is publicly available at https://github.com/FrancoVazza/JULIA/tree/master/ROGER.


\bibliographystyle{mnras}
\bibliography{lib} 




\appendix


\bsp	
\label{lastpage}
\end{document}